\PassOptionsToPackage{unicode}{hyperref}
\PassOptionsToPackage{hyphens}{url}
\documentclass[
]{article}
\usepackage{lmodern}
\usepackage{amssymb,amsmath}
\usepackage{ifxetex,ifluatex}
\ifnum 0\ifxetex 1\fi\ifluatex 1\fi=0 
  \usepackage[T1]{fontenc}
  \usepackage[utf8]{inputenc}
  \usepackage{textcomp} 
\else 
  \usepackage{unicode-math}
  \defaultfontfeatures{Scale=MatchLowercase}
  \defaultfontfeatures[\rmfamily]{Ligatures=TeX,Scale=1}
\fi
\IfFileExists{upquote.sty}{\usepackage{upquote}}{}
\IfFileExists{microtype.sty}{
  \usepackage[]{microtype}
  \UseMicrotypeSet[protrusion]{basicmath} 
}{}
\makeatletter
\@ifundefined{KOMAClassName}{
  \IfFileExists{parskip.sty}{%
    \usepackage{parskip}
  }{
    \setlength{\parindent}{0pt}
    \setlength{\parskip}{6pt plus 2pt minus 1pt}}
}{
  \KOMAoptions{parskip=half}}
\makeatother
\usepackage{xcolor}
\IfFileExists{xurl.sty}{\usepackage{xurl}}{} 
\IfFileExists{bookmark.sty}{\usepackage{bookmark}}{\usepackage{hyperref}}
\hypersetup{
  pdftitle={Current and Next Generation Survey Filter Conversions with ProSpect},
  pdfauthor={Aaron Robotham (aaron.robotham@uwa.edu.au)},
  hidelinks,
  pdfcreator={LaTeX via pandoc}}
\usepackage[margin=1in]{geometry}
\usepackage{color}
\usepackage{fancyvrb}

\DefineVerbatimEnvironment{Highlighting}{Verbatim}{commandchars=\\\{\}}
\usepackage{framed}
\definecolor{shadecolor}{RGB}{248,248,248}
\newenvironment{Shaded}{\begin{snugshade}}{\end{snugshade}}

\newcommand{\DataTypeTok}[1]{\textcolor[rgb]{0.13,0.29,0.53}{#1}}
\newcommand{\DecValTok}[1]{\textcolor[rgb]{0.00,0.00,0.81}{#1}}

\newcommand{\FloatTok}[1]{\textcolor[rgb]{0.00,0.00,0.81}{#1}}

\newcommand{\KeywordTok}[1]{\textcolor[rgb]{0.13,0.29,0.53}{\textbf{#1}}}
\newcommand{\NormalTok}[1]{#1}
\newcommand{\OperatorTok}[1]{\textcolor[rgb]{0.81,0.36,0.00}{\textbf{#1}}}
\newcommand{\OtherTok}[1]{\textcolor[rgb]{0.56,0.35,0.01}{#1}}

\newcommand{\StringTok}[1]{\textcolor[rgb]{0.31,0.60,0.02}{#1}}

\usepackage{longtable,booktabs}
\usepackage{etoolbox}
\makeatletter
\patchcmd\longtable{\par}{\if@noskipsec\mbox{}\fi\par}{}{}
\makeatother
\IfFileExists{footnotehyper.sty}{\usepackage{footnotehyper}}{\usepackage{footnote}}
\makesavenoteenv{longtable}
\usepackage{graphicx,grffile}
\makeatletter
\def\maxwidth{\ifdim\Gin@nat@width>\linewidth\linewidth\else\Gin@nat@width\fi}
\def\maxheight{\ifdim\Gin@nat@height>\textheight\textheight\else\Gin@nat@height\fi}
\makeatother
\setkeys{Gin}{width=\maxwidth,height=\maxheight,keepaspectratio}
\makeatletter
\def\fps@figure{htbp}
\makeatother
\providecommand{\tightlist}{%
  \setlength{\itemsep}{0pt}\setlength{\parskip}{0pt}}
\title{Current and Next Generation Survey Filter Conversions with ProSpect}
\author{Aaron Robotham
(\href{mailto:aaron.robotham@uwa.edu.au}{\nolinkurl{aaron.robotham@uwa.edu.au}})}
\date{27/02/2020}

\begin{document}
\maketitle

{
\setcounter{tocdepth}{2}
\tableofcontents
}
\hypertarget{abstract}{%
\section{Abstract}\label{abstract}}

In this work we compute a reasonably comprehensive set of tables for
current and next generation survey facility filter conversions. Almost
all useful transforms are included with the \textbf{ProSpect} software
package described in Robotham, et al. (2020). Users are free to provide
their own filters and compute their own transforms, where the included
package examples outline the approach. This arXiv document will be
relatively frequently updated, so people are encouraged to get in touch
with their suggestions for additional utility (i.e.~new filter sets).

\hypertarget{introduction}{%
\section{Introduction}\label{introduction}}

Converting between filters from different facilities is an important
activity in astronomy since we are often trying to compare results from
slightly inhomogeneous data sets. Whilst these conversions cannot be
done perfectly, these tables use a physically motivated galaxy formation
model \textbf{Shark} (Lagos, et al. 2018) processed with the \textbf{R}
based \textbf{ProSpect} SED package (Robotham, et al. 2020) applying
sensible dust prescriptions to generate best-effort filter mappings as a
function of redshift. The latter is important compared to many tables
available online, since many of these conversions change significantly
with redshift as certain strong features (e.g.~4,000 Angstrom break)
slide in and out of filters.

Here we focus on major optical and NIR survey facilities (presently, and
soon to come online), where we convert all filters to target Sloan
telescope optical (ugriz) and VISTA NIR (ZYJHKs, though note the Ks
filter is also referred to as K in this document) filters. The filter
responses for these target filters can be seen in the Figure below. The
main current reference for Sloan filter transforms is
\url{http://www.sdss3.org/dr10/algorithms/sdssUBVRITransform.php}, which
explicitly notes that the transforms presented there are not optimised
for galaxies (they are all based on stars and quasars). The most recent
set of transforms for VISTA in the near-infrared can be found in
González-Fernández, et al. (2018).

\includegraphics{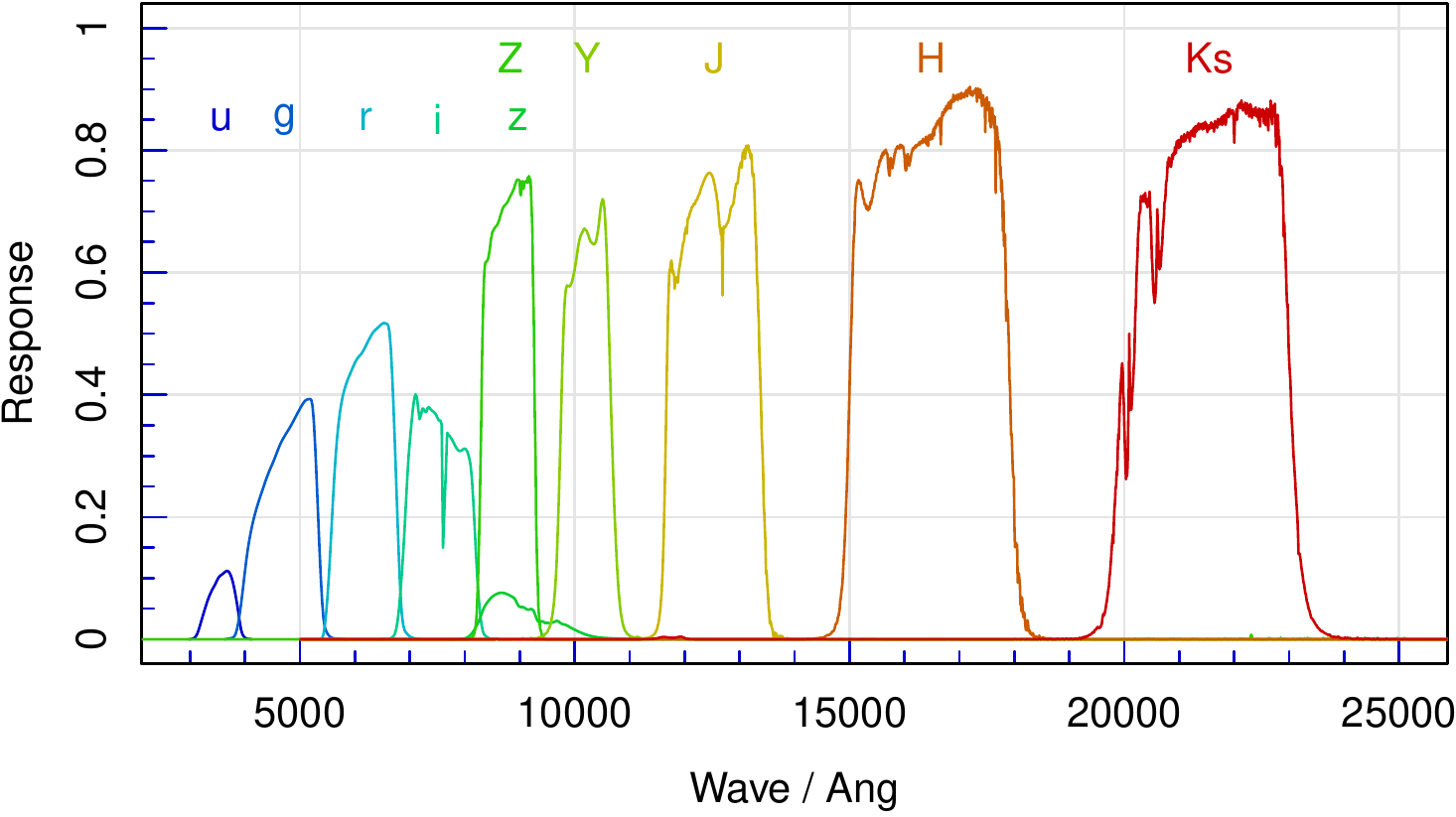}

In this reference document we are interested in creating filter
transforms that work well for galaxies over a reasonable range in
redshift. By combining the \textbf{ProSpect} SED code with the plausible
star formation histories (SFH) and metallicity histories (ZH) produced
by \textbf{Shark}, we hope to capture the dominant transform effects.
For moderate and high redshift galaxies (\(z>0.5\)) these transforms
should be a much more reasonable approximation to the usual sets that
are available in reference papers (and littered around haphazardly on
poorly maintained websites). For a number of current and upcoming
surveys, these transforms appear to be entirely novel.

Finally, users are able to generate their own arbitrary mapping between
any filter sets available in \textbf{ProSpect} using the Shiny web tool
available at \url{http://transformcalc.icrar.org}. Currently it is not
possible to upload a user defined filter to the web tool because of the
computational time required to process the large sample of SFHs (roughly
10-20 seconds for the few 10,000s used). This can be done using the
standalone \textbf{ProSpect} software however. A vignette describing how
to do this is available at \url{https://rpubs.com/asgr/567881}. The
caveat here is that the SFHs sampled are not taken from \textbf{Shark}
(these take up too much memory to include with the \textbf{ProSpect}
package), but are instead randomly generated by sampling the free
parameters of the \emph{massfunc\_b5} star formation history function.
Users can also specify the redshift window that they compute their
transforms over, which might be more accurate than the coarser grid
supplied here.

\hypertarget{methods}{%
\section{Methods}\label{methods}}

Below we outline the linear form of the filter transform used, where we
do not include an explicit redshift dependency term. Instead we compute
each transform for 14 steps of redshift extending from \(z=0\) to
\(z\sim7\), which covers most practical use cases in observational
extra-galactic astronomy.

\hypertarget{filter-mapping-equation}{%
\subsection{Filter Mapping Equation}\label{filter-mapping-equation}}

In all cases (where \(F\) is the Target and Reference filter),
\(\alpha\) is alpha, \(\beta\) is beta and \(\sigma\) is sigma) the
filter mapping is defined by the linear colour equation

\[
F_{\textrm{Tar}} = F_{\textrm{Ref}} + \alpha {\textrm{Col}} + \beta \pm \sigma.
\]

To determine the \emph{best} solution for a given target filter and
reference facility we compute the minimum of

\[
2\alpha^2 + \beta^2 + \sigma^2
\]

for all possible adjacent linear combinations of the reference facility
filters. This solution is the one presented in the following tables, and
is a sensible definition of \emph{best} since it will naturally minimise
the effect of photometric errors when computing the mappings, assuming
these are similar across the filters.

In general \(F_{\textrm{Ref}}\) is the filter that is natively
\emph{closest} to the \(F_{\textrm{Tar}}\) target filter. In the worst
case scenario of only having a single filter pre transform, this should
be used with the specified \(\beta\) part of the equation to achieve an
approximate transform. Obviously, if a user has access to colour
information via having multiple reference filters then a more accurate
transform is possible.

The wary user should probably disregard solutions where the scatter
(\(\sigma\)) is \textgreater{} 0.1. This would suggest a very poor
filter mapping. Also, solutions where \(\alpha\) \textgreater{} 0.5 are
dubious, having a very significant colour term that will enhance any
colour noise present considerably. However, for consistency we keep all
best solutions in these tables (reliable or not). Also, note that a
large value of \(\beta\) (\(>0.1\)) implies that the target and
reference filters are far apart, and the user might want to exercise
caution since the colour terms will become less accurate (e.g.~the
mapping might have become highly non-linear).

In summary, users should choose reference filters as close as possible
to the target filter, and ignore solutions that appear larger than
suggested above for the \(\alpha\), \(\beta\) and \(\sigma\) terms.

\hypertarget{cross-facility-mapping-equation}{%
\subsection{Cross Facility Mapping
Equation}\label{cross-facility-mapping-equation}}

In general, if one wishes to convert between facilities that are not
SDSS or VISTA, the following strategy is appropriate. Take an example
galaxy at \(z=0.417\) mapping to the r\_SDSS filter, where we actually
want to convert between r\_VST and r\_HSC. We find the following
mappings from the tables published below (ignoring the \(\sigma\) term,
that just estimates the error in the conversion):

\[
r_{SDSS} = r_{VST} + 0.050 (g_{VST} - r_{VST}) - 0.010 \\
r_{SDSS} = r_{HSC} - 0.009 (g_{HSC} - r_{HSC}) + 0.016
\]

With a bit of re-arrangement we get to

\[
r_{HSC} = r_{VST} + 0.050 (g_{VST} - r_{VST}) + 0.009 (g_{HSC} - r_{HSC}) - 0.026.
\]

Now we make a reasonable (but approximate) assumption that our colour
terms (being relative) can be used interchangeably. This gets us to

\[
r_{HSC} = r_{VST} + 0.059 (g - r)  - 0.026.
\]

This is almost exactly the same solution that we recover when directly
using \textbf{ProSpect} to compute the \(r_{HSC}\) to \(r_{VST}\)
mapping, suggesting our colour approximation should in general work well
when the filters are similar. As a consequence, cross facility
conversions should use

\[
F_{\textrm{Tar}} = F_{\textrm{Ref}} + (\alpha_{\textrm{Ref}} - \alpha_{\textrm{Tar}}) {\textrm{Col}} + (\beta_{\textrm{Ref}} - \beta_{\textrm{Tar}}).
\]

The pseudo \(\beta\) values will be more accurate than the pseudo
\(\alpha\) since we are not making any assumption on the colour
behaviour being similar. This will only work reasonably when all of the
\(\alpha\), \(\beta\) and \(\sigma\) terms are small (we recommend all
should be less than 0.1).

The scatter (\(\sigma\)) is non-trivial to estimate when mapping across
filter sets, but a pessimistic estimate is to add the reference
\(\sigma\) in quadrature, i.e.:

\[
F_{\textrm{Tar}} = F_{\textrm{Ref}} + (\alpha_{\textrm{Ref}} - \alpha_{\textrm{Tar}}) {\textrm{Col}} + (\beta_{\textrm{Ref}} - \beta_{\textrm{Tar}}) \pm \sqrt{\sigma_{\textrm{Ref}}^2 + \sigma_{\textrm{Tar}}^2}.
\]

As discussed previously, users can convert more directly between filters
that are not the included references sets using the webtool provided at
\url{http://transformcalc.icrar.org}.

\hypertarget{application-of-conversions-to-empirical-data}{%
\section{Application of Conversions To Empirical
Data}\label{application-of-conversions-to-empirical-data}}

Below we add a simple example of trying to create target HSC H band
photometry based on VISTA photometry using \textbf{ProSpect} within an
\textbf{R} session. We wish to apply this to the DEVILS survey (Davies,
et al. 2018)\}, which has a typical redshift of around 0.5, so we use
the 5 Gyr age SFHs to compute the mapping.

\begin{Shaded}
\begin{Highlighting}[]
\NormalTok{tarY_HSC =}\StringTok{ }\KeywordTok{filterTranMags}\NormalTok{(ProFiltTrans_Shark}\OperatorTok{$}\NormalTok{maglist}\OperatorTok{$}\NormalTok{Age5[,}\KeywordTok{c}\NormalTok{(}\StringTok{"Z_VISTA"}\NormalTok{, }\StringTok{"Y_VISTA"}\NormalTok{,}
           \StringTok{"J_VISTA"}\NormalTok{)], ProFiltTrans_Shark}\OperatorTok{$}\NormalTok{maglist}\OperatorTok{$}\NormalTok{Age5[,}\StringTok{"Y_HSC"}\NormalTok{], }\DataTypeTok{return =} \StringTok{'bestall'}\NormalTok{)}
\KeywordTok{print}\NormalTok{(tarY_HSC}\OperatorTok{$}\NormalTok{params)}
\end{Highlighting}
\end{Shaded}

\begin{verbatim}
## $`Y_VISTA + alpha.(Y_VISTA - J_VISTA) + beta +/- sigma`
##        alpha         beta        sigma 
##  0.253892139 -0.013014625  0.003892386
\end{verbatim}

The best solution uses (Y\_VISTA - J\_VISTA) colour data. We can check
how this looks against DEVILS D10 data:

\begin{Shaded}
\begin{Highlighting}[]
\KeywordTok{maghist}\NormalTok{(D10[,mag_HY_t }\OperatorTok{-}\StringTok{ }\NormalTok{mag_Y_t], }\DataTypeTok{breaks=}\KeywordTok{seq}\NormalTok{(}\OperatorTok{-}\DecValTok{1}\NormalTok{,}\DecValTok{1}\NormalTok{,}\DataTypeTok{by=}\FloatTok{0.01}\NormalTok{), }\DataTypeTok{verbose=}\OtherTok{FALSE}\NormalTok{, }\DataTypeTok{xlim=}\KeywordTok{c}\NormalTok{(}\OperatorTok{-}\FloatTok{0.5}\NormalTok{,}\FloatTok{0.5}\NormalTok{),}
        \DataTypeTok{ylim=}\KeywordTok{c}\NormalTok{(}\DecValTok{0}\NormalTok{,}\FloatTok{2e3}\NormalTok{), }\DataTypeTok{grid=}\OtherTok{TRUE}\NormalTok{, }\DataTypeTok{xlab=}\StringTok{'Y_HSC - Y_VISTA/Y_Transform'}\NormalTok{, }\DataTypeTok{ylab=}\StringTok{'Counts'}\NormalTok{)}
\KeywordTok{maghist}\NormalTok{(D10[,mag_HY_t }\OperatorTok{-}\StringTok{ }\NormalTok{(mag_Y_t }\OperatorTok{+}\StringTok{ }\FloatTok{0.2539}\OperatorTok{*}\NormalTok{(mag_Y_t }\OperatorTok{-}\StringTok{ }\NormalTok{mag_J_t) }\OperatorTok{+}\StringTok{ }\FloatTok{-0.0130}\NormalTok{)],}
        \DataTypeTok{breaks=}\KeywordTok{seq}\NormalTok{(}\OperatorTok{-}\DecValTok{1}\NormalTok{,}\DecValTok{1}\NormalTok{,}\DataTypeTok{by=}\FloatTok{0.01}\NormalTok{), }\DataTypeTok{verbose=}\NormalTok{F, }\DataTypeTok{add=}\NormalTok{T, }\DataTypeTok{border=}\StringTok{'red'}\NormalTok{)}
\KeywordTok{legend}\NormalTok{(}\StringTok{'topright'}\NormalTok{, }\DataTypeTok{legend=}\KeywordTok{c}\NormalTok{(}\StringTok{'Y_VISTA'}\NormalTok{,}\StringTok{'Y_Transform'}\NormalTok{), }\DataTypeTok{col=}\KeywordTok{c}\NormalTok{(}\StringTok{'black'}\NormalTok{,}\StringTok{'red'}\NormalTok{), }\DataTypeTok{lty=}\DecValTok{1}\NormalTok{)}
\end{Highlighting}
\end{Shaded}

\includegraphics{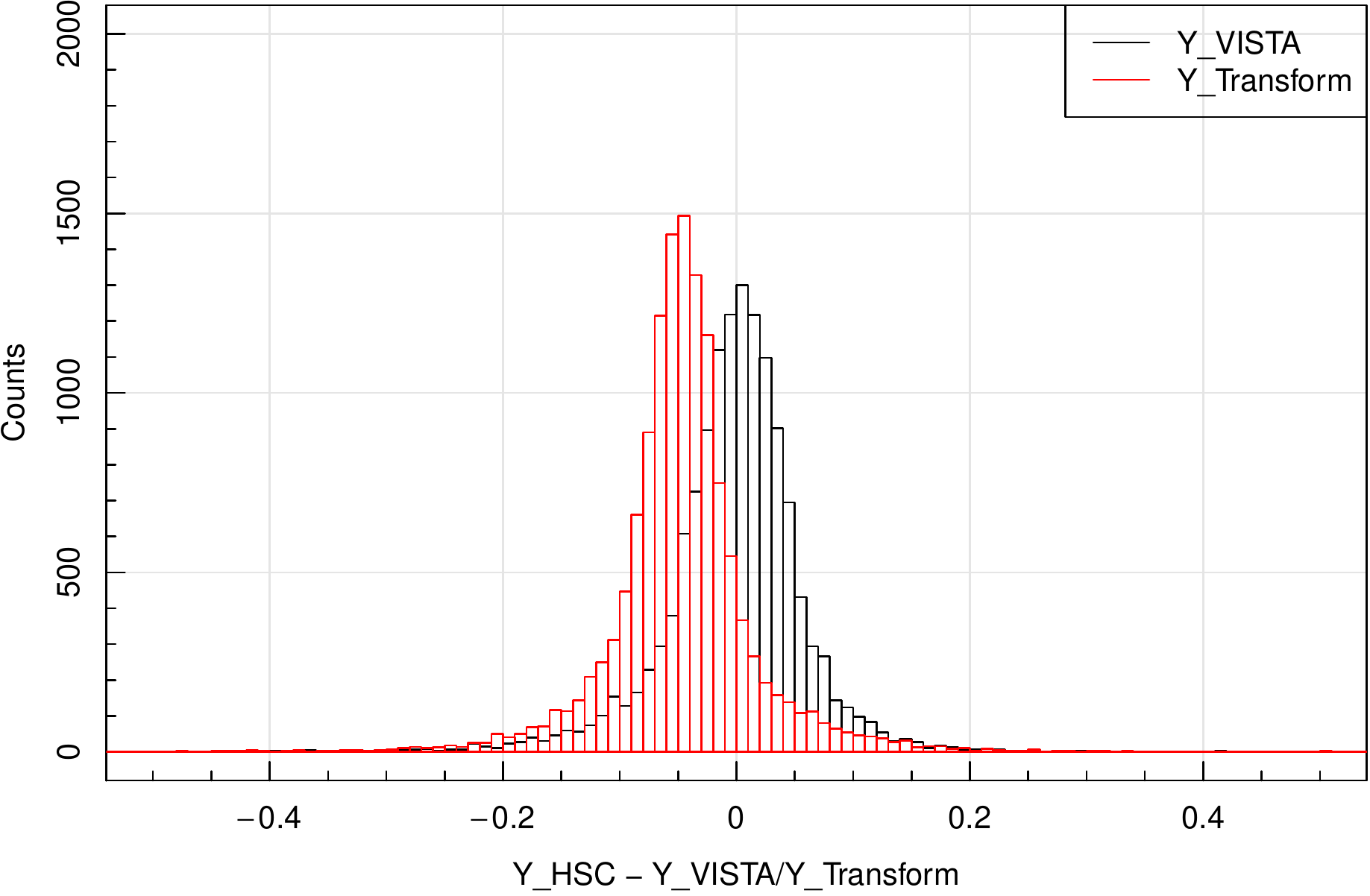}

It is pretty clear we have a zero-point photometry issue here. The
distribution is tighter, but offset from our target of zero. We can
apply the offset and improve the result though:

\begin{Shaded}
\begin{Highlighting}[]
\KeywordTok{maghist}\NormalTok{(D10[,mag_HY_t }\OperatorTok{-}\StringTok{ }\NormalTok{mag_Y_t], }\DataTypeTok{breaks=}\KeywordTok{seq}\NormalTok{(}\OperatorTok{-}\DecValTok{1}\NormalTok{,}\DecValTok{1}\NormalTok{,}\DataTypeTok{by=}\FloatTok{0.01}\NormalTok{), }\DataTypeTok{verbose=}\OtherTok{FALSE}\NormalTok{,}
        \DataTypeTok{xlim=}\KeywordTok{c}\NormalTok{(}\OperatorTok{-}\FloatTok{0.5}\NormalTok{,}\FloatTok{0.5}\NormalTok{), }\DataTypeTok{ylim=}\KeywordTok{c}\NormalTok{(}\DecValTok{0}\NormalTok{,}\FloatTok{2e3}\NormalTok{), }\DataTypeTok{grid=}\OtherTok{TRUE}\NormalTok{, }\DataTypeTok{xlab=}\StringTok{'Y_HSC - Y_VISTA/Y_Transform'}\NormalTok{,}
        \DataTypeTok{ylab=}\StringTok{'Counts'}\NormalTok{)}
\KeywordTok{maghist}\NormalTok{(D10[,mag_HY_t }\OperatorTok{-}\StringTok{ }\NormalTok{(mag_Y_t }\OperatorTok{+}\StringTok{ }\FloatTok{0.2539}\OperatorTok{*}\NormalTok{(mag_Y_t }\OperatorTok{-}\StringTok{ }\NormalTok{mag_J_t) }\OperatorTok{+}\StringTok{ }\FloatTok{-0.0603}\NormalTok{)],}
        \DataTypeTok{breaks=}\KeywordTok{seq}\NormalTok{(}\OperatorTok{-}\DecValTok{1}\NormalTok{,}\DecValTok{1}\NormalTok{,}\DataTypeTok{by=}\FloatTok{0.01}\NormalTok{), }\DataTypeTok{verbose=}\NormalTok{F, }\DataTypeTok{add=}\NormalTok{T, }\DataTypeTok{border=}\StringTok{'red'}\NormalTok{)}
\KeywordTok{legend}\NormalTok{(}\StringTok{'topright'}\NormalTok{, }\DataTypeTok{legend=}\KeywordTok{c}\NormalTok{(}\StringTok{'Y_VISTA'}\NormalTok{,}\StringTok{'Y_Transform'}\NormalTok{), }\DataTypeTok{col=}\KeywordTok{c}\NormalTok{(}\StringTok{'black'}\NormalTok{,}\StringTok{'red'}\NormalTok{), }\DataTypeTok{lty=}\DecValTok{1}\NormalTok{)}
\end{Highlighting}
\end{Shaded}

\includegraphics{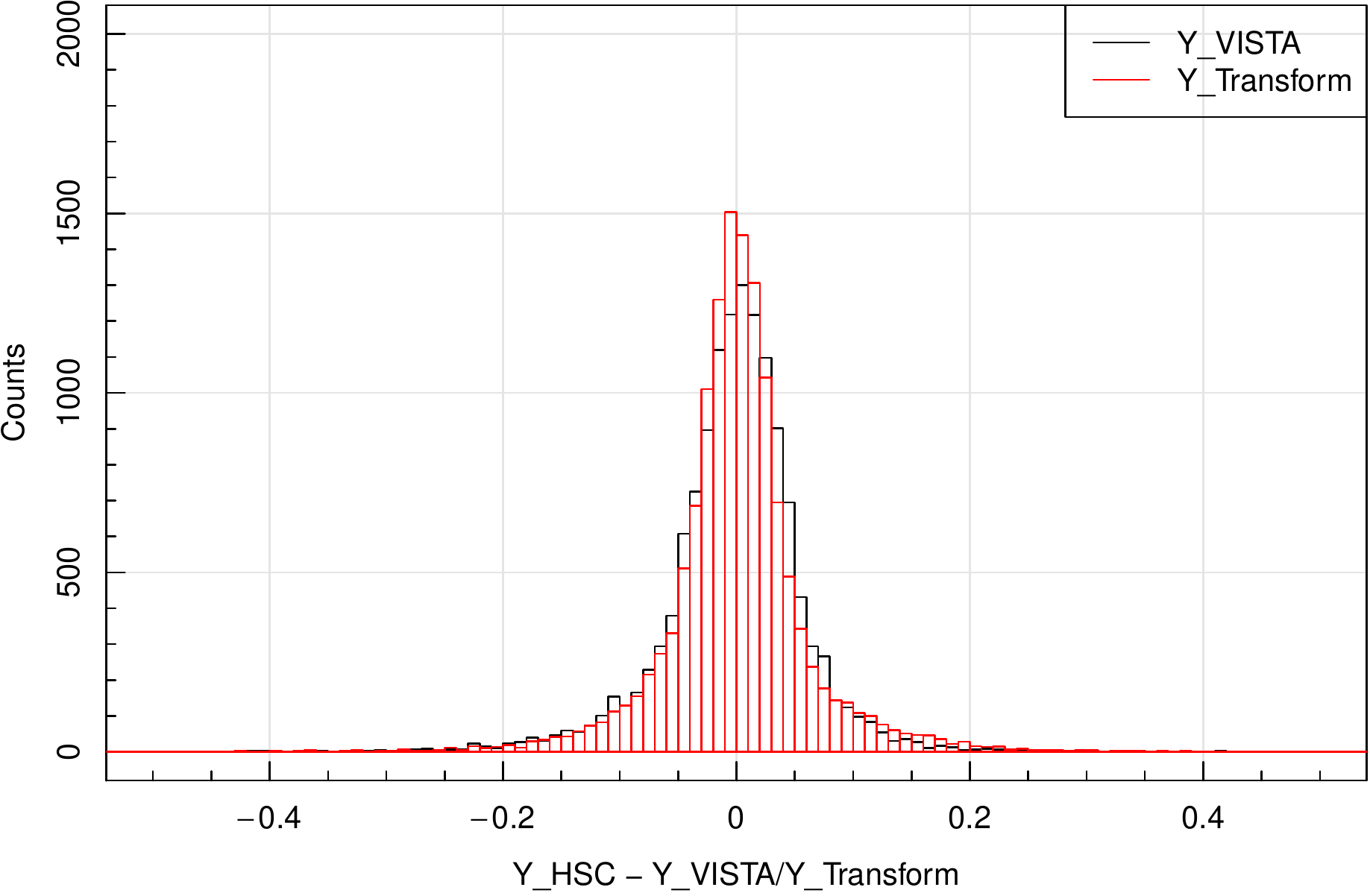}

The above issue highlights that for similar filters the systematic
uncertainties due to zero-points probably often dominate over any
idealised transform equation. In this case we have no way of knowing
whether the error exists in the Y\_HSC or Y\_VISTA photometry, or
perhaps both. The only route to understanding this over all bands is to
analyse the multi-band residuals from full \textbf{ProSpect} SED fits.

We can carry out a test that is less sensitive to such zero-point
issues. Instead of trying to predict small filter changes we can
generate the transforms for a filter that sits inside a large gap
between filters. In this case we will attempt to create a transform for
DEVILS that predicts the Y\_VISTA filter given surrounding Z\_VISTA and
J\_VISTA filters. In this sense \textbf{ProSpect} can be used to create
reasonable missing photometry for galaxies. This might be useful where a
user wishes to create a Y\_VISTA selection cut, but this data happens to
missing for a particular galaxy. If the other photometric data are
considered usable, a user might still want to use the galaxy for
scientific analysis. This approach is clearly \emph{much} cheaper
computationally than trying to do full \textbf{ProSpect} SED fits to the
data and infer the photometry with the outputs (which would also work).
In fact, if spectroscopic or photometric redshift data is not available
it might be more reasonable to just use an approximate transform.

\begin{Shaded}
\begin{Highlighting}[]
\NormalTok{tarY_VISTA =}\StringTok{ }\KeywordTok{filterTranMags}\NormalTok{(ProFiltTrans_Shark}\OperatorTok{$}\NormalTok{maglist}\OperatorTok{$}\NormalTok{Age4[,}\KeywordTok{c}\NormalTok{(}\StringTok{"Z_VISTA"}\NormalTok{, }\StringTok{"J_VISTA"}\NormalTok{)],}
\NormalTok{               ProFiltTrans_Shark}\OperatorTok{$}\NormalTok{maglist}\OperatorTok{$}\NormalTok{Age4[,}\StringTok{"Y_VISTA"}\NormalTok{], }\DataTypeTok{return =} \StringTok{'bestall'}\NormalTok{)}
\KeywordTok{print}\NormalTok{(tarY_VISTA}\OperatorTok{$}\NormalTok{params)}
\end{Highlighting}
\end{Shaded}

\begin{verbatim}
## $`Z_VISTA + alpha.(Z_VISTA - J_VISTA) + beta +/- sigma`
##       alpha        beta       sigma 
## -0.41812772 -0.04485583  0.01526601
\end{verbatim}

\begin{Shaded}
\begin{Highlighting}[]
\KeywordTok{maghist}\NormalTok{(D10[,mag_Y_t }\OperatorTok{-}\StringTok{ }\NormalTok{mag_Z_t], }\DataTypeTok{breaks=}\KeywordTok{seq}\NormalTok{(}\OperatorTok{-}\DecValTok{1}\NormalTok{,}\DecValTok{1}\NormalTok{,}\DataTypeTok{by=}\FloatTok{0.01}\NormalTok{), }\DataTypeTok{verbose=}\OtherTok{FALSE}\NormalTok{,}
        \DataTypeTok{xlim=}\KeywordTok{c}\NormalTok{(}\OperatorTok{-}\FloatTok{0.5}\NormalTok{,}\FloatTok{0.5}\NormalTok{), }\DataTypeTok{ylim=}\KeywordTok{c}\NormalTok{(}\DecValTok{0}\NormalTok{,}\FloatTok{1.5e3}\NormalTok{), }\DataTypeTok{grid=}\OtherTok{TRUE}\NormalTok{, }\DataTypeTok{xlab=}\StringTok{'Y_VISTA - Z_VISTA/Y_Transform'}\NormalTok{,}
        \DataTypeTok{ylab=}\StringTok{'Counts'}\NormalTok{)}
\KeywordTok{maghist}\NormalTok{(D10[,mag_Y_t }\OperatorTok{-}\StringTok{ }\NormalTok{(mag_Z_t }\OperatorTok{+}\StringTok{ }\FloatTok{-0.4181}\OperatorTok{*}\NormalTok{(mag_Z_t }\OperatorTok{-}\StringTok{ }\NormalTok{mag_J_t) }\OperatorTok{+}\StringTok{ }\FloatTok{-0.0449}\NormalTok{)],}
        \DataTypeTok{breaks=}\KeywordTok{seq}\NormalTok{(}\OperatorTok{-}\DecValTok{1}\NormalTok{,}\DecValTok{1}\NormalTok{,}\DataTypeTok{by=}\FloatTok{0.01}\NormalTok{), }\DataTypeTok{verbose=}\NormalTok{F, }\DataTypeTok{add=}\NormalTok{T, }\DataTypeTok{border=}\StringTok{'red'}\NormalTok{)}
\KeywordTok{legend}\NormalTok{(}\StringTok{'topright'}\NormalTok{, }\DataTypeTok{legend=}\KeywordTok{c}\NormalTok{(}\StringTok{'Z_VISTA'}\NormalTok{,}\StringTok{'Y_Transform'}\NormalTok{), }\DataTypeTok{col=}\KeywordTok{c}\NormalTok{(}\StringTok{'black'}\NormalTok{,}\StringTok{'red'}\NormalTok{), }\DataTypeTok{lty=}\DecValTok{1}\NormalTok{)}
\end{Highlighting}
\end{Shaded}

\includegraphics{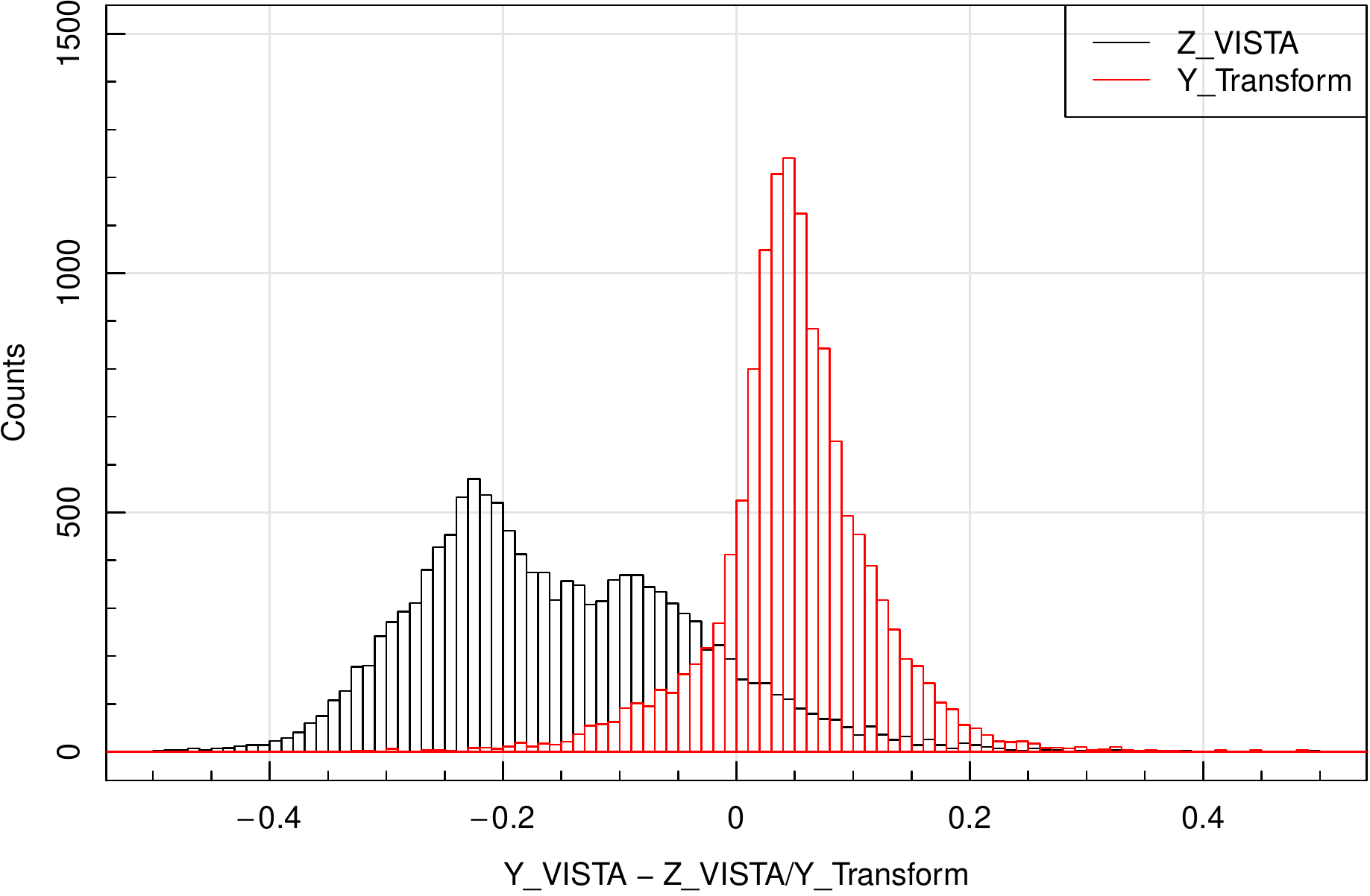}

There is still a small offset from 0 for the red (transformed)
distribution, but it is clearly much tighter and closer to the target
filter.

\hypertarget{conversion-tables}{%
\section{Conversion Tables}\label{conversion-tables}}

Below we list the conversion tables for a number of major current and
upcoming optical and near-infrared survey facilities. This list is not
meant to be exhaustive, as mentioned above specific solutions are
available for many more filters through the \textbf{ProSpect} web tool
(\url{http://transformcalc.icrar.org}), and unavailable filters can be
processed by users using \textbf{ProSpect} locally using approaches
outlined in the online vignette (\url{https://rpubs.com/asgr/567881}).

\hypertarget{vst}{%
\subsection{VST}\label{vst}}

\begin{longtable}[]{@{}lrllrrr@{}}
\caption{Target filter: u\_SDSS, Reference facility: VST}\tabularnewline
\toprule
F\_Tar & z & F\_Ref & Col & alpha & beta & sigma\tabularnewline
\midrule
\endfirsthead
\toprule
F\_Tar & z & F\_Ref & Col & alpha & beta & sigma\tabularnewline
\midrule
\endhead
u\_SDSS & 0.0369 & u & (u - g) & 0.0249 & -0.0298 &
0.0015\tabularnewline
u\_SDSS & 0.1149 & u & (u - g) & 0.0225 & -0.0135 &
0.0008\tabularnewline
u\_SDSS & 0.2025 & u & (u - g) & 0.0098 & -0.0045 &
0.0010\tabularnewline
u\_SDSS & 0.3021 & u & (u - g) & 0.0027 & -0.0004 &
0.0020\tabularnewline
u\_SDSS & 0.4170 & u & (u - g) & 0.0193 & -0.0036 &
0.0012\tabularnewline
u\_SDSS & 0.5519 & u & (u - g) & 0.0058 & 0.0001 & 0.0008\tabularnewline
u\_SDSS & 0.7142 & u & (u - g) & 0.0097 & 0.0015 & 0.0016\tabularnewline
u\_SDSS & 0.9152 & u & (u - g) & 0.0185 & 0.0004 & 0.0008\tabularnewline
u\_SDSS & 1.1746 & u & (u - g) & 0.0044 & -0.0003 &
0.0018\tabularnewline
u\_SDSS & 1.5293 & u & (u - g) & 0.0162 & 0.0056 & 0.0028\tabularnewline
u\_SDSS & 2.0582 & u & (u - g) & -0.0110 & 0.0043 &
0.0026\tabularnewline
u\_SDSS & 2.9779 & u & (u - g) & -0.0564 & 0.0815 &
0.0156\tabularnewline
u\_SDSS & 5.2891 & u & (u - g) & -0.0160 & 0.0460 &
0.0009\tabularnewline
\bottomrule
\end{longtable}

\begin{longtable}[]{@{}lrllrrr@{}}
\caption{Target filter: g\_SDSS, Reference facility: VST}\tabularnewline
\toprule
F\_Tar & z & F\_Ref & Col & alpha & beta & sigma\tabularnewline
\midrule
\endfirsthead
\toprule
F\_Tar & z & F\_Ref & Col & alpha & beta & sigma\tabularnewline
\midrule
\endhead
g\_SDSS & 0.0369 & g & (g - r) & -0.0045 & 0.0061 &
0.0017\tabularnewline
g\_SDSS & 0.1149 & g & (g - r) & 0.0052 & -0.0009 &
0.0010\tabularnewline
g\_SDSS & 0.2025 & g & (u - g) & 0.0133 & -0.0148 &
0.0024\tabularnewline
g\_SDSS & 0.3021 & g & (u - g) & 0.0078 & 0.0064 & 0.0026\tabularnewline
g\_SDSS & 0.4170 & g & (g - r) & 0.0023 & 0.0208 & 0.0027\tabularnewline
g\_SDSS & 0.5519 & g & (u - g) & 0.0040 & 0.0005 & 0.0004\tabularnewline
g\_SDSS & 0.7142 & g & (u - g) & 0.0069 & -0.0003 &
0.0003\tabularnewline
g\_SDSS & 0.9152 & g & (g - r) & -0.0035 & 0.0008 &
0.0008\tabularnewline
g\_SDSS & 1.1746 & g & (u - g) & 0.0084 & -0.0003 &
0.0003\tabularnewline
g\_SDSS & 1.5293 & g & (u - g) & 0.0046 & 0.0006 & 0.0005\tabularnewline
g\_SDSS & 2.0582 & g & (u - g) & 0.0070 & 0.0007 & 0.0014\tabularnewline
g\_SDSS & 2.9779 & g & (u - g) & 0.0173 & -0.0275 &
0.0024\tabularnewline
g\_SDSS & 5.2891 & g & (g - r) & 0.0027 & -0.0017 &
0.0004\tabularnewline
\bottomrule
\end{longtable}

\begin{longtable}[]{@{}lrllrrr@{}}
\caption{Target filter: r\_SDSS, Reference facility: VST}\tabularnewline
\toprule
F\_Tar & z & F\_Ref & Col & alpha & beta & sigma\tabularnewline
\midrule
\endfirsthead
\toprule
F\_Tar & z & F\_Ref & Col & alpha & beta & sigma\tabularnewline
\midrule
\endhead
r\_SDSS & 0.0369 & r & (g - r) & 0.0455 & -0.0056 &
0.0006\tabularnewline
r\_SDSS & 0.1149 & r & (g - r) & 0.0433 & 0.0030 & 0.0014\tabularnewline
r\_SDSS & 0.2025 & r & (g - r) & 0.0521 & -0.0039 &
0.0023\tabularnewline
r\_SDSS & 0.3021 & r & (g - r) & 0.0486 & -0.0117 &
0.0052\tabularnewline
r\_SDSS & 0.4170 & r & (g - r) & 0.0847 & -0.0732 &
0.0065\tabularnewline
r\_SDSS & 0.5519 & r & (g - r) & 0.0784 & -0.0275 &
0.0123\tabularnewline
r\_SDSS & 0.7142 & r & (g - r) & 0.0407 & 0.0589 & 0.0055\tabularnewline
r\_SDSS & 0.9152 & r & (g - r) & 0.0559 & 0.0052 & 0.0031\tabularnewline
r\_SDSS & 1.1746 & r & (g - r) & 0.0957 & -0.0007 &
0.0036\tabularnewline
r\_SDSS & 1.5293 & r & (r - i) & 0.0556 & -0.0003 &
0.0026\tabularnewline
r\_SDSS & 2.0582 & r & (g - r) & 0.0406 & 0.0006 & 0.0021\tabularnewline
r\_SDSS & 2.9779 & r & (g - r) & 0.0384 & -0.0028 &
0.0019\tabularnewline
r\_SDSS & 5.2891 & r & (g - r) & 0.0928 & -0.2914 &
0.0070\tabularnewline
\bottomrule
\end{longtable}

\begin{longtable}[]{@{}lrllrrr@{}}
\caption{Target filter: i\_SDSS, Reference facility: VST}\tabularnewline
\toprule
F\_Tar & z & F\_Ref & Col & alpha & beta & sigma\tabularnewline
\midrule
\endfirsthead
\toprule
F\_Tar & z & F\_Ref & Col & alpha & beta & sigma\tabularnewline
\midrule
\endhead
i\_SDSS & 0.0369 & i & (r - i) & 0.0705 & -0.0094 &
0.0027\tabularnewline
i\_SDSS & 0.1149 & i & (r - i) & 0.0445 & 0.0041 & 0.0020\tabularnewline
i\_SDSS & 0.2025 & i & (r - i) & 0.0329 & 0.0032 & 0.0007\tabularnewline
i\_SDSS & 0.3021 & i & (r - i) & 0.0338 & -0.0016 &
0.0005\tabularnewline
i\_SDSS & 0.4170 & i & (r - i) & 0.0536 & -0.0037 &
0.0017\tabularnewline
i\_SDSS & 0.5519 & i & (r - i) & 0.0295 & 0.0032 & 0.0028\tabularnewline
i\_SDSS & 0.7142 & i & (r - i) & 0.0631 & -0.0519 &
0.0046\tabularnewline
i\_SDSS & 0.9152 & i & (r - i) & 0.0656 & -0.0301 &
0.0071\tabularnewline
i\_SDSS & 1.1746 & i & (r - i) & 0.0338 & 0.0641 & 0.0078\tabularnewline
i\_SDSS & 1.5293 & i & (r - i) & 0.0733 & -0.0014 &
0.0015\tabularnewline
i\_SDSS & 2.0582 & i & (i - z) & 0.0459 & 0.0000 & 0.0006\tabularnewline
i\_SDSS & 2.9779 & i & (r - i) & 0.0415 & 0.0004 & 0.0007\tabularnewline
i\_SDSS & 5.2891 & i & (i - z) & 0.1068 & -0.0014 &
0.0009\tabularnewline
\bottomrule
\end{longtable}

\begin{longtable}[]{@{}lrllrrr@{}}
\caption{Target filter: z\_SDSS, Reference facility: VST}\tabularnewline
\toprule
F\_Tar & z & F\_Ref & Col & alpha & beta & sigma\tabularnewline
\midrule
\endfirsthead
\toprule
F\_Tar & z & F\_Ref & Col & alpha & beta & sigma\tabularnewline
\midrule
\endhead
z\_SDSS & 0.0369 & z & (i - z) & -0.0476 & 0.0024 &
0.0022\tabularnewline
z\_SDSS & 0.1149 & z & (i - z) & -0.0419 & 0.0038 &
0.0029\tabularnewline
z\_SDSS & 0.2025 & z & (i - z) & -0.0530 & 0.0055 &
0.0026\tabularnewline
z\_SDSS & 0.3021 & z & (i - z) & -0.0365 & -0.0042 &
0.0018\tabularnewline
z\_SDSS & 0.4170 & z & (i - z) & -0.0386 & -0.0015 &
0.0010\tabularnewline
z\_SDSS & 0.5519 & z & (i - z) & -0.0228 & 0.0022 &
0.0009\tabularnewline
z\_SDSS & 0.7142 & z & (i - z) & -0.0599 & 0.0110 &
0.0019\tabularnewline
z\_SDSS & 0.9152 & z & (i - z) & -0.0027 & -0.0064 &
0.0036\tabularnewline
z\_SDSS & 1.1746 & z & (i - z) & -0.0834 & 0.0468 &
0.0085\tabularnewline
z\_SDSS & 1.5293 & z & (i - z) & -0.0484 & -0.0651 &
0.0076\tabularnewline
z\_SDSS & 2.0582 & z & (i - z) & -0.0924 & 0.0023 &
0.0017\tabularnewline
z\_SDSS & 2.9779 & z & (i - z) & -0.0662 & 0.0033 &
0.0012\tabularnewline
z\_SDSS & 5.2891 & z & (i - z) & 0.0149 & -0.0003 &
0.0008\tabularnewline
\bottomrule
\end{longtable}

\begin{longtable}[]{@{}lrllrrr@{}}
\caption{Target filter: Z\_VISTA, Reference facility:
VST}\tabularnewline
\toprule
F\_Tar & z & F\_Ref & Col & alpha & beta & sigma\tabularnewline
\midrule
\endfirsthead
\toprule
F\_Tar & z & F\_Ref & Col & alpha & beta & sigma\tabularnewline
\midrule
\endhead
Z\_VISTA & 0.0369 & z & (i - z) & 0.0626 & 0.0030 &
0.0027\tabularnewline
Z\_VISTA & 0.1149 & z & (i - z) & 0.0494 & -0.0024 &
0.0014\tabularnewline
Z\_VISTA & 0.2025 & z & (i - z) & 0.0574 & -0.0023 &
0.0015\tabularnewline
Z\_VISTA & 0.3021 & z & (i - z) & 0.0656 & -0.0016 &
0.0006\tabularnewline
Z\_VISTA & 0.4170 & z & (i - z) & 0.0424 & 0.0076 &
0.0017\tabularnewline
Z\_VISTA & 0.5519 & z & (i - z) & 0.0460 & 0.0022 &
0.0012\tabularnewline
Z\_VISTA & 0.7142 & z & (i - z) & 0.0412 & 0.0195 &
0.0018\tabularnewline
Z\_VISTA & 0.9152 & z & (i - z) & 0.0652 & -0.0169 &
0.0026\tabularnewline
Z\_VISTA & 1.1746 & z & (i - z) & 0.0692 & -0.0427 &
0.0060\tabularnewline
Z\_VISTA & 1.5293 & z & (i - z) & 0.0454 & 0.0537 &
0.0062\tabularnewline
Z\_VISTA & 2.0582 & z & (i - z) & 0.0685 & -0.0013 &
0.0011\tabularnewline
Z\_VISTA & 2.9779 & z & (i - z) & -0.0060 & -0.0014 &
0.0016\tabularnewline
Z\_VISTA & 5.2891 & z & (i - z) & 0.0144 & 0.0017 &
0.0010\tabularnewline
\bottomrule
\end{longtable}

\begin{longtable}[]{@{}lrllrrr@{}}
\caption{Target filter: Y\_VISTA, Reference facility:
VST}\tabularnewline
\toprule
F\_Tar & z & F\_Ref & Col & alpha & beta & sigma\tabularnewline
\midrule
\endfirsthead
\toprule
F\_Tar & z & F\_Ref & Col & alpha & beta & sigma\tabularnewline
\midrule
\endhead
Y\_VISTA & 0.0369 & r & (g - r) & -0.8770 & 0.0143 &
0.0900\tabularnewline
Y\_VISTA & 0.1149 & z & (i - z) & -0.7332 & 0.0304 &
0.0272\tabularnewline
Y\_VISTA & 0.2025 & z & (i - z) & -0.8780 & 0.0676 &
0.0238\tabularnewline
Y\_VISTA & 0.3021 & z & (i - z) & -0.7291 & -0.0163 &
0.0171\tabularnewline
Y\_VISTA & 0.4170 & z & (i - z) & -0.6735 & -0.0105 &
0.0081\tabularnewline
Y\_VISTA & 0.5519 & z & (i - z) & -0.6064 & 0.0193 &
0.0060\tabularnewline
Y\_VISTA & 0.7142 & z & (i - z) & -0.6594 & -0.0492 &
0.0082\tabularnewline
Y\_VISTA & 0.9152 & z & (i - z) & -0.5635 & 0.0623 &
0.0350\tabularnewline
Y\_VISTA & 1.1746 & z & (i - z) & -1.0611 & 0.5462 &
0.0748\tabularnewline
Y\_VISTA & 1.5293 & z & (i - z) & -0.6283 & -0.4970 &
0.0433\tabularnewline
Y\_VISTA & 2.0582 & z & (i - z) & -1.1802 & 0.0066 &
0.0250\tabularnewline
Y\_VISTA & 2.9779 & r & (g - r) & -0.8992 & 0.0671 &
0.0519\tabularnewline
Y\_VISTA & 5.2891 & z & (i - z) & -0.1775 & -0.0031 &
0.0067\tabularnewline
\bottomrule
\end{longtable}

\begin{longtable}[]{@{}lrllrrr@{}}
\caption{Target filter: J\_VISTA, Reference facility:
VST}\tabularnewline
\toprule
F\_Tar & z & F\_Ref & Col & alpha & beta & sigma\tabularnewline
\midrule
\endfirsthead
\toprule
F\_Tar & z & F\_Ref & Col & alpha & beta & sigma\tabularnewline
\midrule
\endhead
J\_VISTA & 0.0369 & r & (g - r) & -1.1451 & 0.1604 &
0.1390\tabularnewline
J\_VISTA & 0.1149 & r & (g - r) & -1.1321 & 0.1517 &
0.1086\tabularnewline
J\_VISTA & 0.2025 & r & (g - r) & -1.1555 & 0.1953 &
0.0634\tabularnewline
J\_VISTA & 0.3021 & g & (u - g) & -1.1864 & -0.3308 &
0.2189\tabularnewline
J\_VISTA & 0.4170 & g & (u - g) & -1.1624 & -0.8231 &
0.2154\tabularnewline
J\_VISTA & 0.5519 & z & (i - z) & -1.3359 & 0.0154 &
0.0273\tabularnewline
J\_VISTA & 0.7142 & r & (g - r) & -1.3198 & -0.0560 &
0.1438\tabularnewline
J\_VISTA & 0.9152 & z & (i - z) & -1.2554 & 0.1242 &
0.0657\tabularnewline
J\_VISTA & 1.1746 & i & (r - i) & -1.8826 & -0.3007 &
0.0829\tabularnewline
J\_VISTA & 1.5293 & z & (i - z) & -1.6377 & -0.3784 &
0.0554\tabularnewline
J\_VISTA & 2.0582 & g & (u - g) & -1.5127 & -0.6486 &
0.3819\tabularnewline
J\_VISTA & 2.9779 & r & (g - r) & -1.5704 & 0.1398 &
0.1073\tabularnewline
J\_VISTA & 5.2891 & z & (i - z) & -0.7883 & 0.0444 &
0.0139\tabularnewline
\bottomrule
\end{longtable}

\begin{longtable}[]{@{}lrllrrr@{}}
\caption{Target filter: H\_VISTA, Reference facility:
VST}\tabularnewline
\toprule
F\_Tar & z & F\_Ref & Col & alpha & beta & sigma\tabularnewline
\midrule
\endfirsthead
\toprule
F\_Tar & z & F\_Ref & Col & alpha & beta & sigma\tabularnewline
\midrule
\endhead
H\_VISTA & 0.0369 & r & (g - r) & -1.4029 & 0.2154 &
0.1927\tabularnewline
H\_VISTA & 0.1149 & r & (g - r) & -1.3756 & 0.2441 &
0.1546\tabularnewline
H\_VISTA & 0.2025 & r & (g - r) & -1.4256 & 0.3819 &
0.1020\tabularnewline
H\_VISTA & 0.3021 & g & (u - g) & -1.3030 & -0.1936 &
0.2872\tabularnewline
H\_VISTA & 0.4170 & g & (u - g) & -1.2813 & -0.7186 &
0.2852\tabularnewline
H\_VISTA & 0.5519 & g & (u - g) & -1.5497 & -1.0195 &
0.3293\tabularnewline
H\_VISTA & 0.7142 & r & (g - r) & -1.5321 & 0.0209 &
0.2053\tabularnewline
H\_VISTA & 0.9152 & r & (g - r) & -1.6011 & -0.6889 &
0.1962\tabularnewline
H\_VISTA & 1.1746 & i & (r - i) & -2.4204 & -0.2569 &
0.1126\tabularnewline
H\_VISTA & 1.5293 & z & (i - z) & -2.4707 & -0.3497 &
0.0743\tabularnewline
H\_VISTA & 2.0582 & g & (u - g) & -1.8526 & -0.6261 &
0.5019\tabularnewline
H\_VISTA & 2.9779 & r & (g - r) & -2.5523 & -0.3032 &
0.2294\tabularnewline
H\_VISTA & 5.2891 & z & (i - z) & -1.2982 & 0.1024 &
0.0280\tabularnewline
\bottomrule
\end{longtable}

\begin{longtable}[]{@{}lrllrrr@{}}
\caption{Target filter: Ks\_VISTA, Reference facility:
VST}\tabularnewline
\toprule
F\_Tar & z & F\_Ref & Col & alpha & beta & sigma\tabularnewline
\midrule
\endfirsthead
\toprule
F\_Tar & z & F\_Ref & Col & alpha & beta & sigma\tabularnewline
\midrule
\endhead
Ks\_VISTA & 0.0369 & r & (g - r) & -1.4129 & 0.4681 &
0.2367\tabularnewline
Ks\_VISTA & 0.1149 & r & (g - r) & -1.4572 & 0.4186 &
0.2043\tabularnewline
Ks\_VISTA & 0.2025 & r & (g - r) & -1.6271 & 0.5288 &
0.1476\tabularnewline
Ks\_VISTA & 0.3021 & g & (u - g) & -1.3813 & -0.1645 &
0.3438\tabularnewline
Ks\_VISTA & 0.4170 & g & (u - g) & -1.3777 & -0.6881 &
0.3398\tabularnewline
Ks\_VISTA & 0.5519 & g & (u - g) & -1.6623 & -0.9425 &
0.3886\tabularnewline
Ks\_VISTA & 0.7142 & r & (g - r) & -1.7462 & 0.2512 &
0.2688\tabularnewline
Ks\_VISTA & 0.9152 & r & (g - r) & -1.8031 & -0.5563 &
0.2514\tabularnewline
Ks\_VISTA & 1.1746 & i & (r - i) & -2.7868 & -0.1661 &
0.1571\tabularnewline
Ks\_VISTA & 1.5293 & z & (i - z) & -2.9876 & -0.2973 &
0.1086\tabularnewline
Ks\_VISTA & 2.0582 & g & (u - g) & -2.0351 & -0.6515 &
0.5727\tabularnewline
Ks\_VISTA & 2.9779 & r & (g - r) & -3.1240 & -0.1923 &
0.3075\tabularnewline
Ks\_VISTA & 5.2891 & z & (i - z) & -2.5664 & 0.2959 &
0.0585\tabularnewline
\bottomrule
\end{longtable}

\hypertarget{hsc}{%
\subsection{HSC}\label{hsc}}

\begin{longtable}[]{@{}lrllrrr@{}}
\caption{Target filter: u\_SDSS, Reference facility: HSC}\tabularnewline
\toprule
F\_Tar & z & F\_Ref & Col & alpha & beta & sigma\tabularnewline
\midrule
\endfirsthead
\toprule
F\_Tar & z & F\_Ref & Col & alpha & beta & sigma\tabularnewline
\midrule
\endhead
u\_SDSS & 0.0369 & g & g & 1.1965 & 0.5920 & 0.0414\tabularnewline
u\_SDSS & 0.1149 & g & g & 1.2119 & 0.5403 & 0.0519\tabularnewline
u\_SDSS & 0.2025 & g & g & 1.3668 & 0.2782 & 0.0956\tabularnewline
u\_SDSS & 0.3021 & g & g & 1.6169 & -0.1734 & 0.1456\tabularnewline
u\_SDSS & 0.4170 & g & g & 2.0605 & -1.0568 & 0.1581\tabularnewline
u\_SDSS & 0.5519 & g & g & 1.7399 & -1.0458 & 0.1399\tabularnewline
u\_SDSS & 0.7142 & g & g & 1.1136 & -0.3634 & 0.0909\tabularnewline
u\_SDSS & 0.9152 & g & g & 0.7442 & -0.0337 & 0.0594\tabularnewline
u\_SDSS & 1.1746 & g & g & 1.1073 & -0.0278 & 0.0712\tabularnewline
u\_SDSS & 1.5293 & g & g & 1.5664 & -0.0223 & 0.0756\tabularnewline
u\_SDSS & 2.0582 & g & g & 1.9675 & 0.0886 & 0.1859\tabularnewline
u\_SDSS & 2.9779 & g & g & 1.5971 & 1.8041 & 0.1103\tabularnewline
u\_SDSS & 5.2891 & g & g & 0.4202 & -0.2251 & 0.0921\tabularnewline
\bottomrule
\end{longtable}

\begin{longtable}[]{@{}lrllrrr@{}}
\caption{Target filter: g\_SDSS, Reference facility: HSC}\tabularnewline
\toprule
F\_Tar & z & F\_Ref & Col & alpha & beta & sigma\tabularnewline
\midrule
\endfirsthead
\toprule
F\_Tar & z & F\_Ref & Col & alpha & beta & sigma\tabularnewline
\midrule
\endhead
g\_SDSS & 0.0369 & g & g & 0.0462 & 0.0028 & 0.0033\tabularnewline
g\_SDSS & 0.1149 & g & g & 0.0734 & -0.0143 & 0.0025\tabularnewline
g\_SDSS & 0.2025 & g & g & 0.0741 & 0.0008 & 0.0043\tabularnewline
g\_SDSS & 0.3021 & g & g & 0.0565 & 0.0333 & 0.0038\tabularnewline
g\_SDSS & 0.4170 & g & g & 0.0320 & 0.0561 & 0.0100\tabularnewline
g\_SDSS & 0.5519 & g & g & 0.0776 & -0.0433 & 0.0042\tabularnewline
g\_SDSS & 0.7142 & g & g & 0.0655 & -0.0200 & 0.0037\tabularnewline
g\_SDSS & 0.9152 & g & g & 0.0280 & -0.0004 & 0.0028\tabularnewline
g\_SDSS & 1.1746 & g & g & 0.0431 & -0.0014 & 0.0013\tabularnewline
g\_SDSS & 1.5293 & g & g & 0.0584 & 0.0009 & 0.0026\tabularnewline
g\_SDSS & 2.0582 & g & g & 0.0671 & -0.0010 & 0.0021\tabularnewline
g\_SDSS & 2.9779 & g & g & 0.1204 & -0.0045 & 0.0044\tabularnewline
g\_SDSS & 5.2891 & g & g & 0.0147 & 0.0013 & 0.0026\tabularnewline
\bottomrule
\end{longtable}

\begin{longtable}[]{@{}lrllrrr@{}}
\caption{Target filter: r\_SDSS, Reference facility: HSC}\tabularnewline
\toprule
F\_Tar & z & F\_Ref & Col & alpha & beta & sigma\tabularnewline
\midrule
\endfirsthead
\toprule
F\_Tar & z & F\_Ref & Col & alpha & beta & sigma\tabularnewline
\midrule
\endhead
r\_SDSS & 0.0369 & r & r & 0.0065 & -0.0076 & 0.0005\tabularnewline
r\_SDSS & 0.1149 & r & r & 0.0091 & -0.0044 & 0.0006\tabularnewline
r\_SDSS & 0.2025 & r & r & 0.0093 & 0.0056 & 0.0017\tabularnewline
r\_SDSS & 0.3021 & r & r & -0.0108 & 0.0093 & 0.0042\tabularnewline
r\_SDSS & 0.4170 & r & r & 0.0214 & -0.0469 & 0.0050\tabularnewline
r\_SDSS & 0.5519 & r & r & 0.0525 & -0.0607 & 0.0108\tabularnewline
r\_SDSS & 0.7142 & r & r & 0.0100 & 0.0273 & 0.0030\tabularnewline
r\_SDSS & 0.9152 & r & r & 0.0095 & 0.0013 & 0.0013\tabularnewline
r\_SDSS & 1.1746 & r & r & 0.0298 & -0.0076 & 0.0022\tabularnewline
r\_SDSS & 1.5293 & r & r & 0.0072 & 0.0005 & 0.0005\tabularnewline
r\_SDSS & 2.0582 & r & r & -0.0002 & 0.0001 & 0.0014\tabularnewline
r\_SDSS & 2.9779 & r & r & 0.0147 & -0.0038 & 0.0010\tabularnewline
r\_SDSS & 5.2891 & r & r & 0.1898 & -0.1419 & 0.0028\tabularnewline
\bottomrule
\end{longtable}

\begin{longtable}[]{@{}lrllrrr@{}}
\caption{Target filter: i\_SDSS, Reference facility: HSC}\tabularnewline
\toprule
F\_Tar & z & F\_Ref & Col & alpha & beta & sigma\tabularnewline
\midrule
\endfirsthead
\toprule
F\_Tar & z & F\_Ref & Col & alpha & beta & sigma\tabularnewline
\midrule
\endhead
i\_SDSS & 0.0369 & i & i & 0.1435 & -0.0097 & 0.0044\tabularnewline
i\_SDSS & 0.1149 & i & i & 0.1197 & 0.0016 & 0.0038\tabularnewline
i\_SDSS & 0.2025 & i & i & 0.0966 & 0.0043 & 0.0008\tabularnewline
i\_SDSS & 0.3021 & i & i & 0.1007 & -0.0050 & 0.0011\tabularnewline
i\_SDSS & 0.4170 & i & i & 0.1007 & 0.0118 & 0.0018\tabularnewline
i\_SDSS & 0.5519 & i & i & 0.1012 & -0.0081 & 0.0059\tabularnewline
i\_SDSS & 0.7142 & i & i & 0.1553 & -0.0924 & 0.0078\tabularnewline
i\_SDSS & 0.9152 & i & i & 0.1260 & -0.0084 & 0.0098\tabularnewline
i\_SDSS & 1.1746 & i & i & 0.0936 & 0.1049 & 0.0135\tabularnewline
i\_SDSS & 1.5293 & i & i & 0.1663 & -0.0003 & 0.0037\tabularnewline
i\_SDSS & 2.0582 & i & i & 0.1056 & -0.0012 & 0.0032\tabularnewline
i\_SDSS & 2.9779 & i & i & 0.1204 & 0.0014 & 0.0016\tabularnewline
i\_SDSS & 5.2891 & i & i & 0.1992 & -0.1133 & 0.0036\tabularnewline
\bottomrule
\end{longtable}

\begin{longtable}[]{@{}lrllrrr@{}}
\caption{Target filter: z\_SDSS, Reference facility: HSC}\tabularnewline
\toprule
F\_Tar & z & F\_Ref & Col & alpha & beta & sigma\tabularnewline
\midrule
\endfirsthead
\toprule
F\_Tar & z & F\_Ref & Col & alpha & beta & sigma\tabularnewline
\midrule
\endhead
z\_SDSS & 0.0369 & z & z & -0.0639 & -0.0016 & 0.0052\tabularnewline
z\_SDSS & 0.1149 & z & z & -0.0362 & 0.0049 & 0.0035\tabularnewline
z\_SDSS & 0.2025 & z & z & -0.0379 & 0.0049 & 0.0016\tabularnewline
z\_SDSS & 0.3021 & z & z & -0.0636 & 0.0120 & 0.0042\tabularnewline
z\_SDSS & 0.4170 & z & z & -0.0247 & -0.0154 & 0.0015\tabularnewline
z\_SDSS & 0.5519 & z & z & -0.0033 & 0.0018 & 0.0020\tabularnewline
z\_SDSS & 0.7142 & z & z & -0.0831 & 0.0158 & 0.0044\tabularnewline
z\_SDSS & 0.9152 & z & z & 0.0085 & -0.0083 & 0.0056\tabularnewline
z\_SDSS & 1.1746 & z & z & -0.0623 & 0.0472 & 0.0087\tabularnewline
z\_SDSS & 1.5293 & z & z & -0.0421 & -0.1190 & 0.0126\tabularnewline
z\_SDSS & 2.0582 & z & z & -0.0380 & 0.0066 & 0.0012\tabularnewline
z\_SDSS & 2.9779 & z & z & -0.0544 & 0.0033 & 0.0013\tabularnewline
z\_SDSS & 5.2891 & z & z & 0.0292 & -0.0017 & 0.0006\tabularnewline
\bottomrule
\end{longtable}

\begin{longtable}[]{@{}lrllrrr@{}}
\caption{Target filter: Z\_VISTA, Reference facility:
HSC}\tabularnewline
\toprule
F\_Tar & z & F\_Ref & Col & alpha & beta & sigma\tabularnewline
\midrule
\endfirsthead
\toprule
F\_Tar & z & F\_Ref & Col & alpha & beta & sigma\tabularnewline
\midrule
\endhead
Z\_VISTA & 0.0369 & z & z & 0.0573 & 0.0008 & 0.0008\tabularnewline
Z\_VISTA & 0.1149 & z & z & 0.0669 & -0.0014 & 0.0018\tabularnewline
Z\_VISTA & 0.2025 & z & z & 0.0864 & -0.0031 & 0.0004\tabularnewline
Z\_VISTA & 0.3021 & z & z & 0.0524 & 0.0163 & 0.0026\tabularnewline
Z\_VISTA & 0.4170 & z & z & 0.0647 & -0.0038 & 0.0020\tabularnewline
Z\_VISTA & 0.5519 & z & z & 0.0745 & 0.0013 & 0.0003\tabularnewline
Z\_VISTA & 0.7142 & z & z & 0.0354 & 0.0243 & 0.0035\tabularnewline
Z\_VISTA & 0.9152 & z & z & 0.0951 & -0.0192 & 0.0022\tabularnewline
Z\_VISTA & 1.1746 & z & z & 0.1099 & -0.0458 & 0.0056\tabularnewline
Z\_VISTA & 1.5293 & z & z & 0.0624 & 0.0049 & 0.0015\tabularnewline
Z\_VISTA & 2.0582 & z & z & 0.1240 & 0.0022 & 0.0018\tabularnewline
Z\_VISTA & 2.9779 & z & z & 0.0155 & -0.0014 & 0.0013\tabularnewline
Z\_VISTA & 5.2891 & z & z & 0.0292 & 0.0003 & 0.0016\tabularnewline
\bottomrule
\end{longtable}

\begin{longtable}[]{@{}lrllrrr@{}}
\caption{Target filter: Y\_VISTA, Reference facility:
HSC}\tabularnewline
\toprule
F\_Tar & z & F\_Ref & Col & alpha & beta & sigma\tabularnewline
\midrule
\endfirsthead
\toprule
F\_Tar & z & F\_Ref & Col & alpha & beta & sigma\tabularnewline
\midrule
\endhead
Y\_VISTA & 0.0369 & Y & Y & -0.4745 & 0.0237 & 0.0050\tabularnewline
Y\_VISTA & 0.1149 & Y & Y & -0.4938 & -0.0110 & 0.0090\tabularnewline
Y\_VISTA & 0.2025 & Y & Y & -0.4747 & 0.0119 & 0.0038\tabularnewline
Y\_VISTA & 0.3021 & Y & Y & -0.4830 & 0.0148 & 0.0021\tabularnewline
Y\_VISTA & 0.4170 & Y & Y & -0.5124 & 0.0156 & 0.0054\tabularnewline
Y\_VISTA & 0.5519 & Y & Y & -0.5020 & 0.0118 & 0.0042\tabularnewline
Y\_VISTA & 0.7142 & Y & Y & -0.3456 & -0.0068 & 0.0026\tabularnewline
Y\_VISTA & 0.9152 & Y & Y & -0.4234 & -0.0324 & 0.0036\tabularnewline
Y\_VISTA & 1.1746 & Y & Y & -0.4327 & 0.0644 & 0.0090\tabularnewline
Y\_VISTA & 1.5293 & Y & Y & -0.5167 & 0.1044 & 0.0255\tabularnewline
Y\_VISTA & 2.0582 & Y & Y & -0.4158 & -0.0005 & 0.0054\tabularnewline
Y\_VISTA & 2.9779 & Y & Y & -0.8900 & -0.0182 & 0.0116\tabularnewline
Y\_VISTA & 5.2891 & z & z & -0.1857 & -0.0038 & 0.0061\tabularnewline
\bottomrule
\end{longtable}

\begin{longtable}[]{@{}lrllrrr@{}}
\caption{Target filter: J\_VISTA, Reference facility:
HSC}\tabularnewline
\toprule
F\_Tar & z & F\_Ref & Col & alpha & beta & sigma\tabularnewline
\midrule
\endfirsthead
\toprule
F\_Tar & z & F\_Ref & Col & alpha & beta & sigma\tabularnewline
\midrule
\endhead
J\_VISTA & 0.0369 & r & r & -1.3073 & 0.1647 & 0.1374\tabularnewline
J\_VISTA & 0.1149 & r & r & -1.2868 & 0.1494 & 0.1091\tabularnewline
J\_VISTA & 0.2025 & r & r & -1.3200 & 0.1986 & 0.0618\tabularnewline
J\_VISTA & 0.3021 & i & i & -1.1511 & -0.0091 & 0.0717\tabularnewline
J\_VISTA & 0.4170 & i & i & -1.1916 & -0.0102 & 0.0406\tabularnewline
J\_VISTA & 0.5519 & i & i & -1.1943 & 0.0471 & 0.0463\tabularnewline
J\_VISTA & 0.7142 & r & r & -1.4778 & -0.1072 & 0.1424\tabularnewline
J\_VISTA & 0.9152 & z & z & -1.2943 & 0.0364 & 0.0683\tabularnewline
J\_VISTA & 1.1746 & i & i & -1.5632 & -0.1976 & 0.0832\tabularnewline
J\_VISTA & 1.5293 & z & z & -1.8533 & -0.5133 & 0.0544\tabularnewline
J\_VISTA & 2.0582 & Y & Y & -2.0797 & -0.6012 & 0.0637\tabularnewline
J\_VISTA & 2.9779 & r & r & -1.7889 & 0.1518 & 0.1044\tabularnewline
J\_VISTA & 5.2891 & z & z & -0.8600 & 0.0445 & 0.0142\tabularnewline
\bottomrule
\end{longtable}

\begin{longtable}[]{@{}lrllrrr@{}}
\caption{Target filter: H\_VISTA, Reference facility:
HSC}\tabularnewline
\toprule
F\_Tar & z & F\_Ref & Col & alpha & beta & sigma\tabularnewline
\midrule
\endfirsthead
\toprule
F\_Tar & z & F\_Ref & Col & alpha & beta & sigma\tabularnewline
\midrule
\endhead
H\_VISTA & 0.0369 & r & r & -1.5935 & 0.2218 & 0.1906\tabularnewline
H\_VISTA & 0.1149 & r & r & -1.5545 & 0.2425 & 0.1553\tabularnewline
H\_VISTA & 0.2025 & r & r & -1.6179 & 0.3840 & 0.0998\tabularnewline
H\_VISTA & 0.3021 & i & i & -1.6158 & 0.1588 & 0.1306\tabularnewline
H\_VISTA & 0.4170 & i & i & -1.6492 & 0.1526 & 0.0981\tabularnewline
H\_VISTA & 0.5519 & i & i & -1.6739 & 0.2015 & 0.0522\tabularnewline
H\_VISTA & 0.7142 & r & r & -1.7100 & -0.0332 & 0.2037\tabularnewline
H\_VISTA & 0.9152 & r & r & -1.7833 & -0.6983 & 0.1996\tabularnewline
H\_VISTA & 1.1746 & i & i & -2.0272 & -0.1366 & 0.1141\tabularnewline
H\_VISTA & 1.5293 & i & i & -2.5376 & -0.5770 & 0.1661\tabularnewline
H\_VISTA & 2.0582 & z & z & -3.8102 & -0.5124 & 0.1512\tabularnewline
H\_VISTA & 2.9779 & r & r & -2.8949 & -0.2819 & 0.2236\tabularnewline
H\_VISTA & 5.2891 & z & z & -1.4271 & 0.1040 & 0.0272\tabularnewline
\bottomrule
\end{longtable}

\begin{longtable}[]{@{}lrllrrr@{}}
\caption{Target filter: Ks\_VISTA, Reference facility:
HSC}\tabularnewline
\toprule
F\_Tar & z & F\_Ref & Col & alpha & beta & sigma\tabularnewline
\midrule
\endfirsthead
\toprule
F\_Tar & z & F\_Ref & Col & alpha & beta & sigma\tabularnewline
\midrule
\endhead
Ks\_VISTA & 0.0369 & r & r & -1.6080 & 0.4760 & 0.2346\tabularnewline
Ks\_VISTA & 0.1149 & r & r & -1.6422 & 0.4163 & 0.2051\tabularnewline
Ks\_VISTA & 0.2025 & r & r & -1.8401 & 0.5302 & 0.1449\tabularnewline
Ks\_VISTA & 0.3021 & i & i & -1.9485 & 0.2146 & 0.1814\tabularnewline
Ks\_VISTA & 0.4170 & i & i & -2.0215 & 0.2306 & 0.1381\tabularnewline
Ks\_VISTA & 0.5519 & i & i & -2.0718 & 0.3816 & 0.0800\tabularnewline
Ks\_VISTA & 0.7142 & r & r & -1.9443 & 0.1942 & 0.2672\tabularnewline
Ks\_VISTA & 0.9152 & r & r & -2.0025 & -0.5660 & 0.2541\tabularnewline
Ks\_VISTA & 1.1746 & i & i & -2.3412 & -0.0351 & 0.1612\tabularnewline
Ks\_VISTA & 1.5293 & i & i & -2.9291 & -0.5590 & 0.2035\tabularnewline
Ks\_VISTA & 2.0582 & r & r & -4.3298 & -0.6012 & 0.3657\tabularnewline
Ks\_VISTA & 2.9779 & r & r & -3.5375 & -0.1659 & 0.3008\tabularnewline
Ks\_VISTA & 5.2891 & z & z & -2.8407 & 0.3013 & 0.0544\tabularnewline
\bottomrule
\end{longtable}

\hypertarget{lsst-rubin-observatory}{%
\subsection{LSST / Rubin Observatory}\label{lsst-rubin-observatory}}

\begin{longtable}[]{@{}lrllrrr@{}}
\caption{Target filter: u\_SDSS, Reference facility:
LSST}\tabularnewline
\toprule
F\_Tar & z & F\_Ref & Col & alpha & beta & sigma\tabularnewline
\midrule
\endfirsthead
\toprule
F\_Tar & z & F\_Ref & Col & alpha & beta & sigma\tabularnewline
\midrule
\endhead
u\_SDSS & 0.0369 & u & (u - g) & 0.0906 & 0.0526 & 0.0141\tabularnewline
u\_SDSS & 0.1149 & u & (u - g) & 0.2019 & -0.1260 &
0.0112\tabularnewline
u\_SDSS & 0.2025 & u & (u - g) & 0.2116 & -0.1149 &
0.0099\tabularnewline
u\_SDSS & 0.3021 & u & (u - g) & 0.1927 & -0.0797 &
0.0107\tabularnewline
u\_SDSS & 0.4170 & u & (u - g) & 0.0813 & -0.0155 &
0.0099\tabularnewline
u\_SDSS & 0.5519 & u & (u - g) & 0.0823 & -0.0075 &
0.0112\tabularnewline
u\_SDSS & 0.7142 & u & (u - g) & 0.0994 & -0.0017 &
0.0040\tabularnewline
u\_SDSS & 0.9152 & u & (u - g) & 0.1417 & -0.0007 &
0.0060\tabularnewline
u\_SDSS & 1.1746 & u & (u - g) & 0.1730 & -0.0065 &
0.0076\tabularnewline
u\_SDSS & 1.5293 & u & (u - g) & 0.1310 & 0.0079 & 0.0106\tabularnewline
u\_SDSS & 2.0582 & u & (u - g) & 0.1064 & 0.0001 & 0.0173\tabularnewline
u\_SDSS & 2.9779 & u & (u - g) & 0.1011 & 0.5350 & 0.0390\tabularnewline
u\_SDSS & 5.2891 & u & (u - g) & 0.0931 & 0.0741 & 0.0035\tabularnewline
\bottomrule
\end{longtable}

\begin{longtable}[]{@{}lrllrrr@{}}
\caption{Target filter: g\_SDSS, Reference facility:
LSST}\tabularnewline
\toprule
F\_Tar & z & F\_Ref & Col & alpha & beta & sigma\tabularnewline
\midrule
\endfirsthead
\toprule
F\_Tar & z & F\_Ref & Col & alpha & beta & sigma\tabularnewline
\midrule
\endhead
g\_SDSS & 0.0369 & g & (u - g) & 0.0180 & -0.0012 &
0.0042\tabularnewline
g\_SDSS & 0.1149 & g & (g - r) & 0.0631 & -0.0258 &
0.0023\tabularnewline
g\_SDSS & 0.2025 & g & (u - g) & 0.0591 & -0.0307 &
0.0096\tabularnewline
g\_SDSS & 0.3021 & g & (u - g) & 0.0295 & 0.0339 & 0.0072\tabularnewline
g\_SDSS & 0.4170 & g & (g - r) & 0.0052 & 0.0693 & 0.0104\tabularnewline
g\_SDSS & 0.5519 & g & (u - g) & 0.0323 & 0.0022 & 0.0019\tabularnewline
g\_SDSS & 0.7142 & g & (u - g) & 0.0526 & 0.0005 & 0.0034\tabularnewline
g\_SDSS & 0.9152 & g & (g - r) & 0.0327 & -0.0015 &
0.0020\tabularnewline
g\_SDSS & 1.1746 & g & (g - r) & 0.0313 & -0.0016 &
0.0012\tabularnewline
g\_SDSS & 1.5293 & g & (u - g) & 0.0291 & 0.0016 & 0.0017\tabularnewline
g\_SDSS & 2.0582 & g & (u - g) & 0.0252 & -0.0012 &
0.0066\tabularnewline
g\_SDSS & 2.9779 & g & (u - g) & 0.0718 & -0.0874 &
0.0072\tabularnewline
g\_SDSS & 5.2891 & g & (g - r) & 0.0145 & -0.0098 &
0.0025\tabularnewline
\bottomrule
\end{longtable}

\begin{longtable}[]{@{}lrllrrr@{}}
\caption{Target filter: r\_SDSS, Reference facility:
LSST}\tabularnewline
\toprule
F\_Tar & z & F\_Ref & Col & alpha & beta & sigma\tabularnewline
\midrule
\endfirsthead
\toprule
F\_Tar & z & F\_Ref & Col & alpha & beta & sigma\tabularnewline
\midrule
\endhead
r\_SDSS & 0.0369 & r & (g - r) & 0.0093 & -0.0082 &
0.0005\tabularnewline
r\_SDSS & 0.1149 & r & (g - r) & 0.0098 & -0.0026 &
0.0004\tabularnewline
r\_SDSS & 0.2025 & r & (g - r) & 0.0103 & 0.0031 & 0.0011\tabularnewline
r\_SDSS & 0.3021 & r & (r - i) & -0.0029 & 0.0062 &
0.0033\tabularnewline
r\_SDSS & 0.4170 & r & (g - r) & 0.0200 & -0.0355 &
0.0035\tabularnewline
r\_SDSS & 0.5519 & r & (g - r) & 0.0379 & -0.0400 &
0.0079\tabularnewline
r\_SDSS & 0.7142 & r & (g - r) & 0.0050 & 0.0183 & 0.0019\tabularnewline
r\_SDSS & 0.9152 & r & (g - r) & 0.0098 & 0.0002 & 0.0011\tabularnewline
r\_SDSS & 1.1746 & r & (g - r) & 0.0291 & -0.0015 &
0.0015\tabularnewline
r\_SDSS & 1.5293 & r & (r - i) & 0.0166 & 0.0004 & 0.0003\tabularnewline
r\_SDSS & 2.0582 & r & (g - r) & 0.0036 & 0.0001 & 0.0009\tabularnewline
r\_SDSS & 2.9779 & r & (g - r) & 0.0124 & -0.0024 &
0.0008\tabularnewline
r\_SDSS & 5.2891 & r & (r - i) & 0.1655 & -0.1038 &
0.0015\tabularnewline
\bottomrule
\end{longtable}

\begin{longtable}[]{@{}lrllrrr@{}}
\caption{Target filter: i\_SDSS, Reference facility:
LSST}\tabularnewline
\toprule
F\_Tar & z & F\_Ref & Col & alpha & beta & sigma\tabularnewline
\midrule
\endfirsthead
\toprule
F\_Tar & z & F\_Ref & Col & alpha & beta & sigma\tabularnewline
\midrule
\endhead
i\_SDSS & 0.0369 & i & (r - i) & 0.0164 & 0.0000 & 4e-04\tabularnewline
i\_SDSS & 0.1149 & i & (r - i) & 0.0120 & 0.0013 & 2e-04\tabularnewline
i\_SDSS & 0.2025 & i & (r - i) & 0.0140 & -0.0022 & 2e-04\tabularnewline
i\_SDSS & 0.3021 & i & (r - i) & 0.0131 & 0.0007 & 2e-04\tabularnewline
i\_SDSS & 0.4170 & i & (r - i) & 0.0126 & 0.0018 & 3e-04\tabularnewline
i\_SDSS & 0.5519 & i & (r - i) & 0.0112 & -0.0018 & 1e-03\tabularnewline
i\_SDSS & 0.7142 & i & (r - i) & 0.0226 & -0.0097 & 1e-03\tabularnewline
i\_SDSS & 0.9152 & i & (r - i) & 0.0137 & 0.0062 & 6e-04\tabularnewline
i\_SDSS & 1.1746 & i & (r - i) & 0.0141 & -0.0026 & 3e-04\tabularnewline
i\_SDSS & 1.5293 & i & (r - i) & 0.0211 & 0.0001 & 5e-04\tabularnewline
i\_SDSS & 2.0582 & i & (i - z) & 0.0166 & -0.0001 & 2e-04\tabularnewline
i\_SDSS & 2.9779 & i & (r - i) & 0.0159 & 0.0001 & 2e-04\tabularnewline
i\_SDSS & 5.2891 & i & (i - z) & 0.0087 & 0.0000 & 1e-03\tabularnewline
\bottomrule
\end{longtable}

\begin{longtable}[]{@{}lrllrrr@{}}
\caption{Target filter: z\_SDSS, Reference facility:
LSST}\tabularnewline
\toprule
F\_Tar & z & F\_Ref & Col & alpha & beta & sigma\tabularnewline
\midrule
\endfirsthead
\toprule
F\_Tar & z & F\_Ref & Col & alpha & beta & sigma\tabularnewline
\midrule
\endhead
z\_SDSS & 0.0369 & z & (i - z) & -0.2064 & 0.0012 &
0.0068\tabularnewline
z\_SDSS & 0.1149 & z & (i - z) & -0.2009 & 0.0127 &
0.0029\tabularnewline
z\_SDSS & 0.2025 & z & (i - z) & -0.2343 & 0.0151 &
0.0043\tabularnewline
z\_SDSS & 0.3021 & z & (i - z) & -0.1959 & -0.0054 &
0.0028\tabularnewline
z\_SDSS & 0.4170 & z & (i - z) & -0.1700 & 0.0014 &
0.0016\tabularnewline
z\_SDSS & 0.5519 & z & (i - z) & -0.1708 & -0.0002 &
0.0012\tabularnewline
z\_SDSS & 0.7142 & z & (i - z) & -0.1612 & -0.0359 &
0.0038\tabularnewline
z\_SDSS & 0.9152 & z & (i - z) & -0.1812 & 0.0427 &
0.0083\tabularnewline
z\_SDSS & 1.1746 & z & (i - z) & -0.2576 & 0.1161 &
0.0225\tabularnewline
z\_SDSS & 1.5293 & z & (i - z) & -0.1865 & -0.1275 &
0.0145\tabularnewline
z\_SDSS & 2.0582 & z & (z - y) & -0.2949 & 0.0017 &
0.0030\tabularnewline
z\_SDSS & 2.9779 & z & (i - z) & -0.1702 & 0.0046 &
0.0040\tabularnewline
z\_SDSS & 5.2891 & z & (i - z) & -0.0582 & -0.0043 &
0.0017\tabularnewline
\bottomrule
\end{longtable}

\begin{longtable}[]{@{}lrllrrr@{}}
\caption{Target filter: Z\_VISTA, Reference facility:
LSST}\tabularnewline
\toprule
F\_Tar & z & F\_Ref & Col & alpha & beta & sigma\tabularnewline
\midrule
\endfirsthead
\toprule
F\_Tar & z & F\_Ref & Col & alpha & beta & sigma\tabularnewline
\midrule
\endhead
Z\_VISTA & 0.0369 & z & (i - z) & -0.0926 & 0.0032 &
0.0017\tabularnewline
Z\_VISTA & 0.1149 & z & (i - z) & -0.1003 & 0.0056 &
0.0010\tabularnewline
Z\_VISTA & 0.2025 & z & (i - z) & -0.1095 & 0.0060 &
0.0026\tabularnewline
Z\_VISTA & 0.3021 & z & (i - z) & -0.0827 & -0.0023 &
0.0005\tabularnewline
Z\_VISTA & 0.4170 & z & (i - z) & -0.0851 & 0.0110 &
0.0009\tabularnewline
Z\_VISTA & 0.5519 & z & (i - z) & -0.0942 & -0.0003 &
0.0021\tabularnewline
Z\_VISTA & 0.7142 & z & (i - z) & -0.0562 & -0.0205 &
0.0027\tabularnewline
Z\_VISTA & 0.9152 & z & (i - z) & -0.1089 & 0.0311 &
0.0030\tabularnewline
Z\_VISTA & 1.1746 & z & (z - y) & -0.1051 & -0.0275 &
0.0030\tabularnewline
Z\_VISTA & 1.5293 & z & (i - z) & -0.0901 & -0.0023 &
0.0008\tabularnewline
Z\_VISTA & 2.0582 & z & (z - y) & -0.1397 & -0.0013 &
0.0026\tabularnewline
Z\_VISTA & 2.9779 & z & (i - z) & -0.1095 & 0.0001 &
0.0018\tabularnewline
Z\_VISTA & 5.2891 & z & (i - z) & -0.0588 & -0.0022 &
0.0006\tabularnewline
\bottomrule
\end{longtable}

\begin{longtable}[]{@{}lrllrrr@{}}
\caption{Target filter: Y\_VISTA, Reference facility:
LSST}\tabularnewline
\toprule
F\_Tar & z & F\_Ref & Col & alpha & beta & sigma\tabularnewline
\midrule
\endfirsthead
\toprule
F\_Tar & z & F\_Ref & Col & alpha & beta & sigma\tabularnewline
\midrule
\endhead
Y\_VISTA & 0.0369 & y & (z - y) & -0.4673 & 0.0259 &
0.0037\tabularnewline
Y\_VISTA & 0.1149 & y & (z - y) & -0.4174 & -0.0085 &
0.0098\tabularnewline
Y\_VISTA & 0.2025 & y & (z - y) & -0.4273 & 0.0141 &
0.0040\tabularnewline
Y\_VISTA & 0.3021 & y & (z - y) & -0.4893 & 0.0248 &
0.0050\tabularnewline
Y\_VISTA & 0.4170 & y & (z - y) & -0.3880 & -0.0171 &
0.0045\tabularnewline
Y\_VISTA & 0.5519 & y & (z - y) & -0.4068 & 0.0141 &
0.0023\tabularnewline
Y\_VISTA & 0.7142 & y & (z - y) & -0.3499 & 0.0056 &
0.0021\tabularnewline
Y\_VISTA & 0.9152 & y & (z - y) & -0.3020 & -0.0635 &
0.0074\tabularnewline
Y\_VISTA & 1.1746 & y & (z - y) & -0.4143 & 0.0873 &
0.0099\tabularnewline
Y\_VISTA & 1.5293 & y & (z - y) & -0.4659 & 0.0383 &
0.0164\tabularnewline
Y\_VISTA & 2.0582 & y & (z - y) & -0.3506 & 0.0028 &
0.0076\tabularnewline
Y\_VISTA & 2.9779 & y & (z - y) & -0.7973 & -0.0148 &
0.0112\tabularnewline
Y\_VISTA & 5.2891 & z & (i - z) & -0.2455 & -0.0081 &
0.0078\tabularnewline
\bottomrule
\end{longtable}

\begin{longtable}[]{@{}lrllrrr@{}}
\caption{Target filter: J\_VISTA, Reference facility:
LSST}\tabularnewline
\toprule
F\_Tar & z & F\_Ref & Col & alpha & beta & sigma\tabularnewline
\midrule
\endfirsthead
\toprule
F\_Tar & z & F\_Ref & Col & alpha & beta & sigma\tabularnewline
\midrule
\endhead
J\_VISTA & 0.0369 & r & (g - r) & -1.2784 & 0.1619 &
0.1365\tabularnewline
J\_VISTA & 0.1149 & r & (g - r) & -1.2805 & 0.1712 &
0.1066\tabularnewline
J\_VISTA & 0.2025 & r & (g - r) & -1.3281 & 0.2225 &
0.0635\tabularnewline
J\_VISTA & 0.3021 & i & (r - i) & -1.3627 & 0.0058 &
0.0739\tabularnewline
J\_VISTA & 0.4170 & g & (u - g) & -1.2309 & -0.7018 &
0.2148\tabularnewline
J\_VISTA & 0.5519 & i & (r - i) & -1.3651 & 0.0330 &
0.0585\tabularnewline
J\_VISTA & 0.7142 & r & (g - r) & -1.4720 & -0.1339 &
0.1449\tabularnewline
J\_VISTA & 0.9152 & y & (z - y) & -1.5121 & -0.1654 &
0.0149\tabularnewline
J\_VISTA & 1.1746 & y & (z - y) & -1.4091 & 0.1113 &
0.0560\tabularnewline
J\_VISTA & 1.5293 & z & (i - z) & -1.8969 & -0.5372 &
0.0605\tabularnewline
J\_VISTA & 2.0582 & y & (z - y) & -1.6195 & -0.5902 &
0.0716\tabularnewline
J\_VISTA & 2.9779 & r & (g - r) & -1.7874 & 0.1626 &
0.1031\tabularnewline
J\_VISTA & 5.2891 & z & (i - z) & -0.8415 & 0.0362 &
0.0152\tabularnewline
\bottomrule
\end{longtable}

\begin{longtable}[]{@{}lrllrrr@{}}
\caption{Target filter: H\_VISTA, Reference facility:
LSST}\tabularnewline
\toprule
F\_Tar & z & F\_Ref & Col & alpha & beta & sigma\tabularnewline
\midrule
\endfirsthead
\toprule
F\_Tar & z & F\_Ref & Col & alpha & beta & sigma\tabularnewline
\midrule
\endhead
H\_VISTA & 0.0369 & r & (g - r) & -1.5597 & 0.2189 &
0.1896\tabularnewline
H\_VISTA & 0.1149 & r & (g - r) & -1.5489 & 0.2694 &
0.1524\tabularnewline
H\_VISTA & 0.2025 & r & (g - r) & -1.6273 & 0.4133 &
0.1018\tabularnewline
H\_VISTA & 0.3021 & g & (u - g) & -1.5194 & -0.0111 &
0.2786\tabularnewline
H\_VISTA & 0.4170 & g & (u - g) & -1.3580 & -0.5892 &
0.2849\tabularnewline
H\_VISTA & 0.5519 & g & (u - g) & -1.6194 & -0.9917 &
0.3154\tabularnewline
H\_VISTA & 0.7142 & r & (g - r) & -1.7014 & -0.0635 &
0.2068\tabularnewline
H\_VISTA & 0.9152 & r & (g - r) & -1.7914 & -0.6990 &
0.1989\tabularnewline
H\_VISTA & 1.1746 & y & (z - y) & -2.2945 & 0.1428 &
0.0899\tabularnewline
H\_VISTA & 1.5293 & z & (i - z) & -2.7932 & -0.5590 &
0.0804\tabularnewline
H\_VISTA & 2.0582 & g & (u - g) & -2.0565 & -0.6112 &
0.4855\tabularnewline
H\_VISTA & 2.9779 & g & (u - g) & -2.2993 & 2.2036 &
0.3526\tabularnewline
H\_VISTA & 5.2891 & z & (i - z) & -1.3380 & 0.0915 &
0.0299\tabularnewline
\bottomrule
\end{longtable}

\begin{longtable}[]{@{}lrllrrr@{}}
\caption{Target filter: Ks\_VISTA, Reference facility:
LSST}\tabularnewline
\toprule
F\_Tar & z & F\_Ref & Col & alpha & beta & sigma\tabularnewline
\midrule
\endfirsthead
\toprule
F\_Tar & z & F\_Ref & Col & alpha & beta & sigma\tabularnewline
\midrule
\endhead
Ks\_VISTA & 0.0369 & r & (g - r) & -1.5752 & 0.4736 &
0.2336\tabularnewline
Ks\_VISTA & 0.1149 & r & (g - r) & -1.6397 & 0.4462 &
0.2019\tabularnewline
Ks\_VISTA & 0.2025 & r & (g - r) & -1.8490 & 0.5628 &
0.1475\tabularnewline
Ks\_VISTA & 0.3021 & g & (u - g) & -1.6133 & 0.0285 &
0.3350\tabularnewline
Ks\_VISTA & 0.4170 & g & (u - g) & -1.4608 & -0.5524 &
0.3395\tabularnewline
Ks\_VISTA & 0.5519 & g & (u - g) & -1.7415 & -0.9116 &
0.3738\tabularnewline
Ks\_VISTA & 0.7142 & r & (g - r) & -1.9329 & 0.1603 &
0.2707\tabularnewline
Ks\_VISTA & 0.9152 & r & (g - r) & -2.0110 & -0.5668 &
0.2538\tabularnewline
Ks\_VISTA & 1.1746 & i & (r - i) & -2.6246 & -0.4359 &
0.1569\tabularnewline
Ks\_VISTA & 1.5293 & z & (i - z) & -3.3562 & -0.5369 &
0.1087\tabularnewline
Ks\_VISTA & 2.0582 & g & (u - g) & -2.2608 & -0.6349 &
0.5556\tabularnewline
Ks\_VISTA & 2.9779 & g & (u - g) & -2.6725 & 2.7210 &
0.4399\tabularnewline
Ks\_VISTA & 5.2891 & z & (i - z) & -2.5736 & 0.2780 &
0.0618\tabularnewline
\bottomrule
\end{longtable}

\hypertarget{ukirt}{%
\subsection{UKIRT}\label{ukirt}}

\begin{longtable}[]{@{}lrllrrr@{}}
\caption{Target filter: u\_SDSS, Reference facility:
UKIRT}\tabularnewline
\toprule
F\_Tar & z & F\_Ref & Col & alpha & beta & sigma\tabularnewline
\midrule
\endfirsthead
\toprule
F\_Tar & z & F\_Ref & Col & alpha & beta & sigma\tabularnewline
\midrule
\endhead
u\_SDSS & 0.0369 & Z & (Z - Y) & 3.6527 & 1.6000 & 0.3316\tabularnewline
u\_SDSS & 0.1149 & H & (H - K) & 4.7195 & 2.8144 & 0.5250\tabularnewline
u\_SDSS & 0.2025 & Y & (Y - J) & 5.4409 & 1.8612 & 0.5022\tabularnewline
u\_SDSS & 0.3021 & Y & (Y - J) & 5.8717 & 1.7622 & 0.5322\tabularnewline
u\_SDSS & 0.4170 & J & (J - H) & 6.3523 & 2.4594 & 0.6025\tabularnewline
u\_SDSS & 0.5519 & J & (J - H) & 6.4291 & 2.0348 & 0.6660\tabularnewline
u\_SDSS & 0.7142 & J & (J - H) & 7.5205 & 1.4954 & 0.5983\tabularnewline
u\_SDSS & 0.9152 & Z & (Z - Y) & 7.0061 & 0.8097 & 0.2728\tabularnewline
u\_SDSS & 1.1746 & Z & (Z - Y) & 4.2352 & 0.9571 & 0.3382\tabularnewline
u\_SDSS & 1.5293 & Y & (Y - J) & 4.5175 & 1.0251 & 0.3368\tabularnewline
u\_SDSS & 2.0582 & Z & (Z - Y) & 3.9681 & 0.1829 & 0.2938\tabularnewline
u\_SDSS & 2.9779 & J & (J - H) & 2.9697 & 0.6927 & 0.3590\tabularnewline
u\_SDSS & 5.2891 & H & (H - K) & 4.1115 & 6.8442 & 0.2328\tabularnewline
\bottomrule
\end{longtable}

\begin{longtable}[]{@{}lrllrrr@{}}
\caption{Target filter: g\_SDSS, Reference facility:
UKIRT}\tabularnewline
\toprule
F\_Tar & z & F\_Ref & Col & alpha & beta & sigma\tabularnewline
\midrule
\endfirsthead
\toprule
F\_Tar & z & F\_Ref & Col & alpha & beta & sigma\tabularnewline
\midrule
\endhead
g\_SDSS & 0.0369 & Z & (Z - Y) & 2.2815 & 0.6229 & 0.2025\tabularnewline
g\_SDSS & 0.1149 & Z & (Z - Y) & 3.6202 & 0.5635 & 0.2145\tabularnewline
g\_SDSS & 0.2025 & Y & (Y - J) & 3.9299 & 0.9192 & 0.2771\tabularnewline
g\_SDSS & 0.3021 & Y & (Y - J) & 4.1882 & 0.9963 & 0.2651\tabularnewline
g\_SDSS & 0.4170 & J & (J - H) & 4.2646 & 1.7358 & 0.2924\tabularnewline
g\_SDSS & 0.5519 & J & (J - H) & 4.6087 & 1.5911 & 0.3716\tabularnewline
g\_SDSS & 0.7142 & Z & (Z - Y) & 5.8668 & 0.4358 & 0.1258\tabularnewline
g\_SDSS & 0.9152 & Z & (Z - Y) & 5.2834 & 0.7740 & 0.1704\tabularnewline
g\_SDSS & 1.1746 & Z & (Z - Y) & 3.0226 & 0.8727 & 0.2243\tabularnewline
g\_SDSS & 1.5293 & Y & (Y - J) & 3.4450 & 0.9187 & 0.2170\tabularnewline
g\_SDSS & 2.0582 & Z & (Z - Y) & 2.2056 & 0.0471 & 0.1055\tabularnewline
g\_SDSS & 2.9779 & J & (J - H) & 2.0565 & -0.8556 &
0.2046\tabularnewline
g\_SDSS & 5.2891 & H & (H - K) & 3.5761 & 5.3042 & 0.1448\tabularnewline
\bottomrule
\end{longtable}

\begin{longtable}[]{@{}lrllrrr@{}}
\caption{Target filter: r\_SDSS, Reference facility:
UKIRT}\tabularnewline
\toprule
F\_Tar & z & F\_Ref & Col & alpha & beta & sigma\tabularnewline
\midrule
\endfirsthead
\toprule
F\_Tar & z & F\_Ref & Col & alpha & beta & sigma\tabularnewline
\midrule
\endhead
r\_SDSS & 0.0369 & Z & (Z - Y) & 1.2826 & 0.2454 & 0.0796\tabularnewline
r\_SDSS & 0.1149 & Z & (Z - Y) & 1.6846 & 0.2370 & 0.0885\tabularnewline
r\_SDSS & 0.2025 & Y & (Y - J) & 1.9665 & 0.4156 & 0.1414\tabularnewline
r\_SDSS & 0.3021 & Y & (Y - J) & 2.0287 & 0.3973 & 0.1565\tabularnewline
r\_SDSS & 0.4170 & J & (J - H) & 2.8847 & 0.7114 & 0.1888\tabularnewline
r\_SDSS & 0.5519 & J & (J - H) & 3.3855 & 0.6862 & 0.2152\tabularnewline
r\_SDSS & 0.7142 & Z & (Z - Y) & 2.9173 & 0.3008 & 0.0543\tabularnewline
r\_SDSS & 0.9152 & Z & (Z - Y) & 2.6051 & 0.7041 & 0.0859\tabularnewline
r\_SDSS & 1.1746 & Z & (Z - Y) & 1.8372 & 0.7695 & 0.1304\tabularnewline
r\_SDSS & 1.5293 & Z & (Z - Y) & 2.3274 & -1.3224 &
0.1568\tabularnewline
r\_SDSS & 2.0582 & Z & (Z - Y) & 1.2349 & 0.0194 & 0.0440\tabularnewline
r\_SDSS & 2.9779 & Y & (Y - J) & 1.3944 & 0.0414 & 0.0328\tabularnewline
r\_SDSS & 5.2891 & H & (H - K) & 2.2087 & 0.8811 & 0.0809\tabularnewline
\bottomrule
\end{longtable}

\begin{longtable}[]{@{}lrllrrr@{}}
\caption{Target filter: i\_SDSS, Reference facility:
UKIRT}\tabularnewline
\toprule
F\_Tar & z & F\_Ref & Col & alpha & beta & sigma\tabularnewline
\midrule
\endfirsthead
\toprule
F\_Tar & z & F\_Ref & Col & alpha & beta & sigma\tabularnewline
\midrule
\endhead
i\_SDSS & 0.0369 & Z & (Z - Y) & 0.8068 & 0.0598 & 0.0289\tabularnewline
i\_SDSS & 0.1149 & Z & (Z - Y) & 0.9055 & 0.0817 & 0.0294\tabularnewline
i\_SDSS & 0.2025 & Z & (Z - Y) & 0.8420 & 0.0995 & 0.0235\tabularnewline
i\_SDSS & 0.3021 & Z & (Z - Y) & 1.0124 & 0.0141 & 0.0234\tabularnewline
i\_SDSS & 0.4170 & Z & (Z - Y) & 1.3925 & -0.0390 &
0.0106\tabularnewline
i\_SDSS & 0.5519 & Z & (Z - Y) & 1.4987 & 0.0238 & 0.0104\tabularnewline
i\_SDSS & 0.7142 & Z & (Z - Y) & 1.4112 & -0.1241 &
0.0156\tabularnewline
i\_SDSS & 0.9152 & Z & (Z - Y) & 1.3082 & 0.1553 & 0.0598\tabularnewline
i\_SDSS & 1.1746 & Z & (Z - Y) & 0.5947 & 0.6404 & 0.0713\tabularnewline
i\_SDSS & 1.5293 & Z & (Z - Y) & 1.1921 & -0.6830 &
0.0695\tabularnewline
i\_SDSS & 2.0582 & Z & (Z - Y) & 0.7233 & 0.0111 & 0.0208\tabularnewline
i\_SDSS & 2.9779 & Z & (Z - Y) & 0.8173 & 0.0301 & 0.0138\tabularnewline
i\_SDSS & 5.2891 & H & (H - K) & 1.5489 & 0.2217 & 0.0492\tabularnewline
\bottomrule
\end{longtable}

\begin{longtable}[]{@{}lrllrrr@{}}
\caption{Target filter: z\_SDSS, Reference facility:
UKIRT}\tabularnewline
\toprule
F\_Tar & z & F\_Ref & Col & alpha & beta & sigma\tabularnewline
\midrule
\endfirsthead
\toprule
F\_Tar & z & F\_Ref & Col & alpha & beta & sigma\tabularnewline
\midrule
\endhead
z\_SDSS & 0.0369 & Z & (Z - Y) & -0.1088 & -0.0039 &
0.0018\tabularnewline
z\_SDSS & 0.1149 & Z & (Z - Y) & -0.0868 & 0.0024 &
0.0008\tabularnewline
z\_SDSS & 0.2025 & Z & (Z - Y) & -0.0879 & -0.0001 &
0.0011\tabularnewline
z\_SDSS & 0.3021 & Z & (Z - Y) & -0.1119 & -0.0001 &
0.0008\tabularnewline
z\_SDSS & 0.4170 & Z & (Z - Y) & -0.0900 & -0.0097 &
0.0009\tabularnewline
z\_SDSS & 0.5519 & Z & (Z - Y) & -0.0839 & -0.0002 &
0.0016\tabularnewline
z\_SDSS & 0.7142 & Z & (Z - Y) & -0.1340 & 0.0090 &
0.0019\tabularnewline
z\_SDSS & 0.9152 & Z & (Z - Y) & -0.0841 & -0.0009 &
0.0025\tabularnewline
z\_SDSS & 1.1746 & Z & (Z - Y) & -0.1275 & 0.0143 &
0.0043\tabularnewline
z\_SDSS & 1.5293 & Z & (Z - Y) & -0.1438 & -0.0355 &
0.0073\tabularnewline
z\_SDSS & 2.0582 & Z & (Z - Y) & -0.1210 & 0.0024 &
0.0022\tabularnewline
z\_SDSS & 2.9779 & Z & (Z - Y) & -0.0775 & 0.0028 &
0.0013\tabularnewline
z\_SDSS & 5.2891 & Z & (Z - Y) & -0.0020 & -0.0001 &
0.0017\tabularnewline
\bottomrule
\end{longtable}

\begin{longtable}[]{@{}lrllrrr@{}}
\caption{Target filter: Z\_VISTA, Reference facility:
UKIRT}\tabularnewline
\toprule
F\_Tar & z & F\_Ref & Col & alpha & beta & sigma\tabularnewline
\midrule
\endfirsthead
\toprule
F\_Tar & z & F\_Ref & Col & alpha & beta & sigma\tabularnewline
\midrule
\endhead
Z\_VISTA & 0.0369 & Z & (Z - Y) & 0.0126 & 0.0015 &
0.0003\tabularnewline
Z\_VISTA & 0.1149 & Z & (Z - Y) & 0.0115 & 0.0013 &
0.0008\tabularnewline
Z\_VISTA & 0.2025 & Z & (Z - Y) & 0.0158 & 0.0014 &
0.0005\tabularnewline
Z\_VISTA & 0.3021 & Z & (Z - Y) & 0.0100 & 0.0013 &
0.0008\tabularnewline
Z\_VISTA & 0.4170 & Z & (Z - Y) & 0.0159 & -0.0023 &
0.0007\tabularnewline
Z\_VISTA & 0.5519 & Z & (Z - Y) & 0.0187 & 0.0013 &
0.0004\tabularnewline
Z\_VISTA & 0.7142 & Z & (Z - Y) & 0.0046 & 0.0088 &
0.0011\tabularnewline
Z\_VISTA & 0.9152 & Z & (Z - Y) & 0.0286 & -0.0044 &
0.0009\tabularnewline
Z\_VISTA & 1.1746 & Z & (Z - Y) & 0.0149 & 0.0048 &
0.0008\tabularnewline
Z\_VISTA & 1.5293 & Z & (Z - Y) & 0.0064 & -0.0035 &
0.0007\tabularnewline
Z\_VISTA & 2.0582 & Z & (Z - Y) & 0.0005 & 0.0000 &
0.0008\tabularnewline
Z\_VISTA & 2.9779 & Z & (Z - Y) & -0.0218 & -0.0006 &
0.0008\tabularnewline
Z\_VISTA & 5.2891 & Z & (Z - Y) & 0.0208 & 0.0010 &
0.0002\tabularnewline
\bottomrule
\end{longtable}

\begin{longtable}[]{@{}lrllrrr@{}}
\caption{Target filter: Y\_VISTA, Reference facility:
UKIRT}\tabularnewline
\toprule
F\_Tar & z & F\_Ref & Col & alpha & beta & sigma\tabularnewline
\midrule
\endfirsthead
\toprule
F\_Tar & z & F\_Ref & Col & alpha & beta & sigma\tabularnewline
\midrule
\endhead
Y\_VISTA & 0.0369 & Y & (Z - Y) & 0.0698 & -0.0045 &
0.0006\tabularnewline
Y\_VISTA & 0.1149 & Y & (Y - J) & 0.0509 & 0.0053 &
0.0007\tabularnewline
Y\_VISTA & 0.2025 & Y & (Y - J) & 0.0536 & 0.0015 &
0.0009\tabularnewline
Y\_VISTA & 0.3021 & Y & (Y - J) & 0.0458 & 0.0010 &
0.0014\tabularnewline
Y\_VISTA & 0.4170 & Y & (Y - J) & 0.0459 & 0.0019 &
0.0011\tabularnewline
Y\_VISTA & 0.5519 & Y & (Y - J) & 0.0448 & 0.0045 &
0.0007\tabularnewline
Y\_VISTA & 0.7142 & Y & (Z - Y) & 0.0403 & 0.0008 &
0.0008\tabularnewline
Y\_VISTA & 0.9152 & Y & (Z - Y) & 0.0594 & 0.0016 &
0.0008\tabularnewline
Y\_VISTA & 1.1746 & Y & (Z - Y) & 0.0436 & -0.0076 &
0.0018\tabularnewline
Y\_VISTA & 1.5293 & Y & (Y - J) & 0.0541 & 0.0124 &
0.0017\tabularnewline
Y\_VISTA & 2.0582 & Y & (Z - Y) & 0.0478 & -0.0004 &
0.0010\tabularnewline
Y\_VISTA & 2.9779 & Y & (Y - J) & 0.0263 & -0.0002 &
0.0006\tabularnewline
Y\_VISTA & 5.2891 & Y & (Y - J) & 0.0384 & 0.0004 &
0.0005\tabularnewline
\bottomrule
\end{longtable}

\begin{longtable}[]{@{}lrllrrr@{}}
\caption{Target filter: J\_VISTA, Reference facility:
UKIRT}\tabularnewline
\toprule
F\_Tar & z & F\_Ref & Col & alpha & beta & sigma\tabularnewline
\midrule
\endfirsthead
\toprule
F\_Tar & z & F\_Ref & Col & alpha & beta & sigma\tabularnewline
\midrule
\endhead
J\_VISTA & 0.0369 & J & (J - H) & -0.0123 & 0.0001 &
4e-04\tabularnewline
J\_VISTA & 0.1149 & J & (Y - J) & -0.0169 & 0.0004 &
3e-04\tabularnewline
J\_VISTA & 0.2025 & J & (Y - J) & -0.0178 & 0.0030 &
6e-04\tabularnewline
J\_VISTA & 0.3021 & J & (J - H) & -0.0176 & 0.0002 &
5e-04\tabularnewline
J\_VISTA & 0.4170 & J & (J - H) & -0.0138 & -0.0013 &
3e-04\tabularnewline
J\_VISTA & 0.5519 & J & (J - H) & -0.0089 & 0.0004 &
4e-04\tabularnewline
J\_VISTA & 0.7142 & J & (J - H) & -0.0116 & -0.0002 &
5e-04\tabularnewline
J\_VISTA & 0.9152 & J & (J - H) & -0.0131 & -0.0022 &
3e-04\tabularnewline
J\_VISTA & 1.1746 & J & (Y - J) & -0.0112 & -0.0004 &
3e-04\tabularnewline
J\_VISTA & 1.5293 & J & (Y - J) & -0.0113 & -0.0014 &
9e-04\tabularnewline
J\_VISTA & 2.0582 & J & (J - H) & -0.0145 & 0.0007 &
9e-04\tabularnewline
J\_VISTA & 2.9779 & J & (J - H) & -0.0111 & 0.0062 &
4e-04\tabularnewline
J\_VISTA & 5.2891 & J & (Y - J) & -0.0174 & -0.0007 &
2e-04\tabularnewline
\bottomrule
\end{longtable}

\begin{longtable}[]{@{}lrllrrr@{}}
\caption{Target filter: H\_VISTA, Reference facility:
UKIRT}\tabularnewline
\toprule
F\_Tar & z & F\_Ref & Col & alpha & beta & sigma\tabularnewline
\midrule
\endfirsthead
\toprule
F\_Tar & z & F\_Ref & Col & alpha & beta & sigma\tabularnewline
\midrule
\endhead
H\_VISTA & 0.0369 & H & (J - H) & -0.0426 & 0.0017 &
0.0011\tabularnewline
H\_VISTA & 0.1149 & H & (H - K) & -0.0293 & -0.0072 &
0.0010\tabularnewline
H\_VISTA & 0.2025 & H & (H - K) & -0.0155 & -0.0016 &
0.0009\tabularnewline
H\_VISTA & 0.3021 & H & (J - H) & -0.0239 & -0.0021 &
0.0006\tabularnewline
H\_VISTA & 0.4170 & H & (J - H) & -0.0248 & 0.0014 &
0.0003\tabularnewline
H\_VISTA & 0.5519 & H & (J - H) & -0.0267 & 0.0038 &
0.0005\tabularnewline
H\_VISTA & 0.7142 & H & (H - K) & -0.0242 & -0.0021 &
0.0007\tabularnewline
H\_VISTA & 0.9152 & H & (J - H) & -0.0222 & 0.0001 &
0.0009\tabularnewline
H\_VISTA & 1.1746 & H & (J - H) & -0.0256 & 0.0017 &
0.0007\tabularnewline
H\_VISTA & 1.5293 & H & (J - H) & -0.0209 & -0.0002 &
0.0002\tabularnewline
H\_VISTA & 2.0582 & H & (J - H) & -0.0214 & -0.0058 &
0.0009\tabularnewline
H\_VISTA & 2.9779 & H & (J - H) & -0.0251 & 0.0044 &
0.0018\tabularnewline
H\_VISTA & 5.2891 & H & (H - K) & -0.0129 & 0.0005 &
0.0002\tabularnewline
\bottomrule
\end{longtable}%

\hypertarget{mass}{%
\subsection{2MASS}\label{mass}}

%

\hypertarget{jwst}{%
\subsection{JWST}\label{jwst}}

%

\hypertarget{euclid}{%
\subsection{Euclid}\label{euclid}}

%

\hypertarget{wfirst}{%
\subsection{WFIRST}\label{wfirst}}

%

\hypertarget{conclusions}{%
\section{Conclusions}\label{conclusions}}

This work offers a hopefully useful set of filter conversion references,
and will be regularly modified going forwards as new facilities and
filter sets are defined. The main limitations are:

\begin{itemize}
\tightlist
\item
  The physicality of the \textbf{ProSpect} SED processing approach (see
  Robotham, et al. 2020 for details on the implementation),
\item
  The physicality of the galaxies generated by \textbf{Shark}, where we
  are limited by the realism of the star formation and metallicity
  histories generated (see Lagos, et al. 2018 for details on the
  implementation),
\item
  The accuracy of the filters used, where in reality many filters
  degrade and change throughput characteristics over time, and the
  effective throughput is modified by variable water vapour in the
  atmosphere at the time of any given observation etc.
\end{itemize}

Users are able to create arbitrary conversions with a larger set of
filters through our web tool available at
\url{http://transformcalc.icrar.org}. Furthermore, users can specify
their own filters directly, but in this case it is necessary to use
\textbf{ProSpect} directly as per \url{https://rpubs.com/asgr/567881}.

\hypertarget{acknowledgements}{%
\section{Acknowledgements}\label{acknowledgements}}

This document was compiled from a self-generating \textbf{Rmarkdown}
file and processed with \textbf{pandoc}. All figures were created with
the \textbf{magicaxis} \textbf{R} package (Robotham 2016). Thanks to
Sabine Bellstedt for reading over the final document, and Simon Driver
and Luke Davies for useful feedback and testing whilst this document was
being assembled.

\hypertarget{references}{%
\section*{References}\label{references}}
\addcontentsline{toc}{section}{References}

\hypertarget{refs}{}
\leavevmode\hypertarget{ref-DEVILS}{}%
Davies, et al. 2018. \emph{MNRAS} 480: 768.

\leavevmode\hypertarget{ref-VISTA}{}%
González-Fernández, et al. 2018. \emph{MNRAS} 474: 5459.

\leavevmode\hypertarget{ref-Shark}{}%
Lagos, et al. 2018. \emph{MNRAS} 481: 3573.

\leavevmode\hypertarget{ref-magicaxis}{}%
Robotham. 2016. \emph{ASCL} 1604.004.

\leavevmode\hypertarget{ref-ProSpect}{}%
Robotham, et al. 2020. \emph{MNRAS} 495: 905.

\end{document}